\documentclass[aps,prd,twocolumn,groupedaddress,asmsymb,amsmath,amssymb,longbibliography]{revtex4-1}

\usepackage{graphicx}
\usepackage{calligra}
\usepackage{upgreek}
\usepackage{color}
\usepackage{caption}
\usepackage{stackrel}
\usepackage[usenames,dvipsnames]{xcolor}

\numberwithin{equation}{section}

\DeclareMathAlphabet{\mathpzc}{OT1}{pzc}{m}{it}

\newcommand{\be}{\begin{equation}}
\newcommand{\ee}{\end{equation}}
\newcommand{\bea}{\begin{eqnarray}}
\newcommand{\eea}{\end{eqnarray}}
\newcommand{\lb}{\label}

\newcommand{\om}{\omega}
\newcommand{\bj}{{\mbox{\boldmath $\dot{\j}$}}}

\newcommand{\bg}{{\bf g}}
\newcommand{\bv}{{\bf v}}
\newcommand{\bu}{{\bf u}}

\newcommand{\bn}{{\bf n}}

\newcommand{\bt}{{\bf t}}
\newcommand{\bx}{{\bf x}}

\newcommand{\br}{{\bf r}}
\newcommand{\bA}{{\bf A}}

\newcommand{\bI}{{\bf I}}

\newcommand{\bP}{{\bf P}}

\newcommand{\bS}{{\bf S}}
\newcommand{\bV}{{\bf V}}

\newcommand{\beLL}{{\mbox{\boldmath $\ell$}}}

\newcommand{\btheta}{{\mbox{\boldmath $\theta$}}}
\newcommand{\barphi}{{\mbox{\boldmath $\varphi$}}}
\newcommand{\bpsi}{{\mbox{\boldmath $\psi$}}}
\newcommand{\bomega}{{\mbox{\boldmath $\omega$}}}
\newcommand{\bSigma}{{\mbox{\boldmath $\Sigma$}}}
\newcommand{\bsigma}{{\mbox{\boldmath $\sigma$}}}
\newcommand{\btau}{{\mbox{\boldmath $\tau$}}}
\newcommand{\grad}{{\mbox{\boldmath $\nabla$}}}
\newcommand{\bdot}{{\mbox{\boldmath $\cdot$}}}
\newcommand{\bdots}{{\mbox{\boldmath $:$}}}
\newcommand{\bzed}{{\mbox{\boldmath $0$}}}
\newcommand{\Bell}{{\mbox{\boldmath $\ell$}}}
\newcommand{\btimes}{{\mbox{\boldmath $\times$}}}

\begin{document}


\title{The Josephson-Anderson Relation and the Classical D'Alembert Paradox}


\author{Gregory L. Eyink${\,\!}^{1,2}$}
\affiliation{${\,\!}^1$Department of Applied Mathematics \& Statistics, The Johns Hopkins University, Baltimore, MD, USA, 21218}
\affiliation{${\,\!}^2$Department of Physics \& Astronomy, The Johns Hopkins University, Baltimore, MD, USA, 21218}


\date{\today}

\begin{abstract}
Generalizing prior work of P.~W. Anderson and E.~R. Huggins, we show that a  ``detailed Josephson-Anderson
relation'' holds for drag on a finite body held at rest in a classical incompressible fluid flowing with velocity ${\bf V}.$
The relation asserts an exact equality between the instantaneous power consumption by the drag, $-{\bf F}\bdot\bV,$
and the vorticity flux across the potential mass current, $-(1/2)\int dJ\int \epsilon_{ijk}\Sigma_{ij}\,d\ell_k.$ Here
$\Sigma_{ij}$ is the flux in the $i$th coordinate direction of the conserved $j$th component of vorticity and 
the line-integrals over $\beLL$ are taken along streamlines of the potential flow solution $\bu_\phi=\grad\phi$ of the ideal 
Euler equation, carrying mass flux $dJ=\rho\,\bu_\phi\bdot d\bA.$ The results generalize the theories of 
M.~J. Lighthill for flow past a body and, in particular, the steady-state relation $(1/2)\epsilon_{ijk}\langle\Sigma_{jk}\rangle
=\partial_i\langle h\rangle,$ where $h=p+(1/2)|\bu|^2$ is the generalized enthalpy or total pressure, extends 
Lighthill's theory of vorticity generation at solid walls into the interior of the flow. We use these results to explain 
drag on the body in terms of vortex dynamics, unifying the theories for classical fluids and for quantum superfluids.  
The results offer a new solution to the ``D'Alembert paradox'' at infinite Reynolds numbers and imply the necessary 
conditions for turbulent drag reduction.  
\end{abstract}

\pacs{?????}

\maketitle


\section{Introduction}\label{sec:I}

The origin of the Josephson-Anderson relation lies in the work of Josephson 
on tunneling of Cooper pairs through normal-superconductor metal junctions and, in particular, his 
AC effect \cite{josephson1962possible}. However, the relation assumed its most elegant 
and powerful statement in the seminal paper of Anderson on flow in superfluid ${\,\!}^4$He
\cite{anderson1966considerations}. The relation has had sufficient sustained importance 
that it garnered two extensive reviews, thirty years \cite{packard1998role} and fifty years 
\cite{varoquaux2015anderson} after Anderson's original work. In its most basic form, it 
relates the time-derivative of the phase $\theta$ of the superfluid macroscopic wavefunction 
and the chemical potential 
$\mu$ as 
\be \hbar \frac{d\theta}{dt} = -\left(\mu + \frac{1}{2}m |\bu|^2\right) \lb{JArelation} \ee 
where $\bu=\hbar\grad\theta/m$ is the superfluid velocity.  A special significance holds 
for this relation because topological phase-defects exist in superfluids as discrete vortex lines
with circulation quantized in units of $h/m$ \cite{onsager1949,feynman1955chapter}. 
The importance of the Josephson-Anderson relation arises from the intimate connection it 
reveals between force balance and vortex motion. This connection can already 
be understood by applying a space-gradient to Eq.\eqref{JArelation} which yields the equation of 
motion for the superfluid velocity
\be  \frac{d\bu}{dt}=\partial_t\bu + (\bu\bdot\grad)\bu = -(1/m)\grad \mu \ee 
A superfluid would be expected 
to flow without any applied chemical potential gradient (or pressure-gradient, for the 
incompressible isothermal limit), but Anderson showed that a drop in chemical potential 
would occur if there were a time-average flux of quantized vortices across the streamlines 
of the potential superfluid flow \cite{anderson1966considerations}. See also Josephson 
\cite{josephson1965potential} for the corresponding effect of voltage drop in superconductors. 

A significant extension of these ideas was obtained subsequently by Huggins \cite{huggins1970energy} 
whose ``detailed Josephson equation'' was further able to relate vortex motion to 
energy dissipation. See also \cite{zimmermann1993energy,greiter2005electromagnetic,
varoquaux2015anderson} for alternative derivations of Huggins' result. 
Such vortex motion by ``$2\pi$-phase slips'' has been for many years the 
standard paradigm for energy dissipation in low-temperature superfluids and superconductors
\cite{bishop1993resistance}. Furthermore, because Huggins' detailed relation was derived 
without any sort of averaging, it could be applied to individual 
flow realizations and it was found to yield deep insight into otherwise very complicated vortex 
dynamics. We may quote from the review of Packard, who wrote that 
\begin{quotation} 
\noindent 
``the equation provides an elegant short cut to certain predictions ...
that involve the 
complex motion of quantized vortices. These same predictions by other methods 
require a detailed knowledge of vortex motion and involve considerable 
computational effort.'' \cite{packard1998role}
\end{quotation} 
The reviews \cite{packard1998role,varoquaux2015anderson} both analyzed a large number 
of concrete examples to justify the remarkable efficacy of the detailed Josephson-Anderson relation 
to understand and predict complex superfluid vortex dynamics. 

That classical fluids should satisfy a similar relation was already understood by Anderson,
who devoted Appendix B of his 1966 article to deriving an analogous result for classical hydrodynamics
\cite{anderson1966considerations}. He stated that he had been unable to find the same result written 
anywhere in the literature and conjectured that ``it was understood by the `classics' but is of no value 
in classical hydrodynamics so was never stated'' \footnote{The author had a conversation with Anderson
about this during a sabbatical at Princeton in 2017 and Anderson still maintained the opinion at that time 
that the relation was of special importance in superfluids where vortices are quantized and was not obviously 
useful in classical hydrodynamics}. Most superfluid physicists have followed Anderson in discounting 
any significance of the Josephson-Anderson relation for classical fluids. As a typical statement,
we may quote from the recent review article of  Varoquaux: 
\begin{quotation}
\noindent 
``This result is of no special importance in classical hydrodynamics because the 
velocity circulation carried by each vortex, albeit constant, can take any value, 
while in the superfluid it is directly related to the phase of the macroscopic wave 
function and quantized.'' \cite{varoquaux2015anderson}\\ 
\end{quotation} 

\vspace{-15pt}\noindent 
A notable exception to this trend was Huggins, whose original paper \cite{huggins1970energy} 
already demonstrated the validity of his ``detailed relation'' for classical viscous Navier-Stokes solutions 
and who later wrote a paper applying his results to classical turbulent channel flow \cite{huggins1994vortex} 
\footnote{Elisha Huggins was a PhD student of Richard Feynman, from whom he may have inherited his interest 
in classical fluid turbulence}. 

Unrecognized by Anderson and the rest of the superfluid community, however, there were 
already important applications of closely related ideas in classical hydrodynamics.
A very early foreshadowing was work on turbulent pipe flow by Taylor \cite{taylor1932transport},
who realized that pressure-drop down the pipe implies transverse vortex motion but who 
did not pursue the connection further.  More important was a seminal work of Lighthill \cite{lighthill1963introduction},  
who presented a very broad vision of classical fluid mechanics from the perspective of vortex dynamics, 
encompassing laminar, transitional and fully turbulent  flow. Just three years before Anderson's work 
on superfluids, Lighthill argued that vorticity flux from solid walls is fundamentally due to tangential 
pressure gradients at the wall and he presented many important applications of this principle to classical
incompressible fluid dynamics. In fact, his ideas are closely related to those of Huggins
\cite{huggins1970exact,huggins1970energy,huggins1971dynamical,huggins1994vortex}, 
which naturally extend Lighthill's concept of vorticity flux at solid walls into the 
the interior of the flow. These connections were previously discussed by us in a paper on turbulent 
channel flow \cite{eyink2008turbulent}, which further developed Huggins ideas on that problem 
and put them in the context of contemporary work on the ``attached-eddy hypothesis''. 

In this paper we present a new application of the classical Josephson-Anderson relation 
to flow past a finite, solid body and to the problem of the origin of drag. This 
is in some ways a much more illuminating application of the relation, although it requires 
some significant modification of the analysis both of Huggins and of Lighthill. 
In fact,  Lighthill in his paper \cite{lighthill1963introduction} had discussed 
this same problem in the rest frame of the fluid, appealing there to Kelvin's 
minimum-energy theorem \cite{kelvin1849vis}.  We shall see that this theorem does 
not hold in the body frame, and this fact crucially enters into the analysis.  
Kelvin's theorem involves the unique potential flow solution of the inviscid Euler equation 
satisfying the no-flow-through condition at the body surface, which, according to the 
famous result of d'Alembert predicts zero drag around a moving body 
\cite{dalembert1749theoria,dalembert1768paradoxe,grimberg2008genesis}. 
In superfluids this is no ``paradox'' but is instead a physically observable phenomenon when the body 
moves at low speeds below a critical velocity. The origin of superfluid drag above this critical velocity 
was the subject of a study by Frisch, Pomeau and Rica \cite{frisch1992transition},
which spawned a substantial following literature up to the present time, 
e.g. \cite{jackson1998vortex,winiecki1999pressure,winiecki2000vortex,winiecki2000motion,
winiecki2001numerical,huepe2000scaling,nore2000subcritical,sasaki2010benard,
stagg2014quantum,stagg2015generation,musser2019starting}.  
Our analysis will demonstrate a deep similarity between the origin of drag  in classical and 
quantum fluids, with the Josephson-Anderson relation providing the key unifying concept. 
Although we shall focus mainly on the classical case, we shall make various remarks 
as we proceed concerning differences and similarities with the quantum case. 
Our approach applies to classical flows at any Reynolds number but in particular extends 
to the infinite-Reynolds turbulent regime, making a connection of the D'Alembert paradox 
with the Onsager theory of dissipative Euler solutions \cite{onsager1949statistical,
eyink2006onsager,eyink2018review}.  

\section{Concise Review of Theories of Huggins and Lighthill}\lb{sec:review} 

We begin with a very brief review of the essential ideas in the theories
of Huggins \cite{huggins1970exact,huggins1970energy,huggins1971dynamical,huggins1994vortex}
and Lighthill  \cite{lighthill1963introduction}. 

Although we shall be principally concerned with simple fluids described by the incompressible
Navier-Stokes equation at constant density $\rho,$ the ideas of Huggins apply to an extended
system of equations with additional accelerations, due to forces both conservative $-\grad U$ 
for a potential $U$ and non-conservative $\bg$ with $\grad\btimes\bg\neq \bzed$, written in the form: 
\be \partial_t\bu=\bu\btimes\bomega-\nu\grad\btimes\bomega -\bg
-\grad\left(p+U+\frac{1}{2}|\bu|^2\right), \lb{NS-ext} \ee 
where $p=P/\rho$ is kinematic pressure and $\nu=\eta/\rho$ is kinematic viscosity.  For example,
$U$ could be a gravitational or electrostatic potential and $\bg$ could arise from the stress 
of a polymer additive. Huggins \cite{huggins1971dynamical,huggins1994vortex} 
observed that equations of the above class can be rewritten component-wise as 
\be \partial_t u_i = \frac{1}{2}\epsilon_{ijk} \Sigma_{jk}  -\partial_i h \lb{NS-ext2} \ee 
in terms of an anti-symmetric tensor 
\be \Sigma_{ij}=u_i\omega_j-u_j\omega_i
+\nu\left(\frac{\partial\omega_i}{\partial x_j}-\frac{\partial\omega_j}{\partial x_i}\right)
-\epsilon_{ijk} g_k \lb{Sigma} \ee 
and a generalized enthalpy or total pressure (static pressure plus dynamic pressure) 
\be h=p+U+\frac{1}{2}|\bu|^2. \ee 
The meaning of the tensor $\bSigma$ is discovered by taking the curl of equation \eqref{NS-ext} 
to obtain the analogue of the Helmholtz equation for conservation of vorticity
\be \partial_t\bomega+\grad\bdot\bSigma=\bzed \lb{vort-cons} \ee 
with $\Sigma_{ij}$ representing the flux of the $j$th component of vorticity in the $i$th coordinate 
direction. For this reason, we shall refer to $\bSigma$ as the {\it Huggins vorticity-flux tensor}. 

The various terms in Eq.\eqref{Sigma} have transparent physical meanings, with the 
first representing advective transport, the second transport by vortex-stretching/tilting, the third 
term in parentheses viscous transport and the final term a flux by the Magnus effect 
transverse to the applied force. The rewriting of the momentum conservation equation 
in the form \eqref{NS-ext2} is the classical Josephson-Anderson relation in its simplest version.
It implies, for example, that for a steady solution or for a suitable time-average $\langle\cdot\rangle$
(average over a period for an oscillatory solution or long-time ergodic average for a chaotic solution)
the mean gradients of $h$ and the mean vorticity fluxes are exactly related by 
\be \frac{1}{2}\epsilon_{ijk} \langle\Sigma_{jk}\rangle =\partial_i \langle h\rangle.  \lb{gradh-Sigma} \ee 
Thus, a mean gradient of $h$ must always be associated to a transverse vorticity flux, and vice versa.   

Furthermore, Huggins \cite{huggins1970exact,huggins1970energy} 
(and see also \cite{eyink2008turbulent}, Appendix B) derived a less obvious 
result, the ``detailed Josephson relation'', in the case of a generalized channel flow, as 
pictured in Figure \ref{fig1}. Here the fluid velocity is assumed to be specified
on the in-flow surface $S_{in},$ out-flow surface $S_{out},$  and at the channel 
wall $S_w$ as 
\be \left.\bu\right|_{S_{in}}=\bu_{in}, \quad \left.\bu\right|_{S_{out}}=\bu_{out}, \quad 
\left.\bu\right|_{S_{w}}=\bzed \ee
As in the proof of the Kelvin minimum energy theorem (\cite{kelvin1849vis}; 
\cite{lamb1945hydrodynamics}, Art.45; \cite{batchelor2000introduction},\S 6.2; 
\cite{wu2007vorticity}, \S 2.4.4), Huggins
then introduced the unique incompressible potential flow $\bu_\phi=\grad\phi$ satisfying 
the same boundary conditions:
\be \left.\bu_\phi\right|_{S_{in}}=\bu_{in}, \quad \left.\bu_\phi\right|_{S_{out}}=\bu_{out}, \quad 
\left.\bu_\phi\right|_{S_{w}}=\bzed \ee 
and the complementary field $\bu_\omega=\bu-\bu_\phi$ which represents the velocity due to 
vorticity. It then easily follows that $\bu_\phi$ and $\bu_\omega$ are orthogonal
\be \int_\Omega \bu_\phi\bdot\bu_\omega\, dV= 
      \int_\Omega \grad\cdot(\phi\bu_\omega)\, dV=\int_{\partial\Omega}\phi (\bu_\omega\bdot\hat{\bn})\,dA=0\ee 
which is the essence of Kelvin's theorem. Using this orthogonality, Huggins was able to derive
equations for energies in the potential flow $E_\phi=(\rho/2)\int_\Omega |\bu_\phi|^2 dV$ and in 
the rotational flow $E_\omega=(\rho/2)\int_\Omega |\bu_\omega|^2 dV$ as
\be \frac{dE_\phi}{dt} = \int dJ (h_{in}-h_{out}) - {\mathcal T} \lb{Ephi-H}  \ee
and
\be \frac{dE_\omega}{dt} = {\mathcal T} -\int_\Omega [\eta|\bomega|^2+\rho\bu\bdot\bg] \, dV
\lb{Eomega-H} \ee
with
\begin{eqnarray} 
{\mathcal T} &=& -\rho \int \bu_\phi\bdot(\bu\btimes\bomega-\nu\grad\btimes\bomega-\bg)\, dV \cr 
&=& -\int dJ \int (\bu\btimes\bomega-\nu\grad\btimes\bomega-\bg)\bdot \,d\Bell \cr 
&= &-\frac{1}{2} \int dJ \int \epsilon_{ijk} \Sigma_{ij} \, d\ell_k  \cr
&&
\end{eqnarray} 
Here the line integrals are along streamlines of the potential flow and $dJ=\rho\bu_\phi\bdot d\bA$ is 
the element of mass flux along each streamline, with ${\mathcal T}$ representing transfer of energy 
from potential to rotational flow by flux of vorticity across mass current. As a consequence 
of Eq.\eqref{Ephi-H}, Huggins then obtained the ``detailed Josephson relation'' 
\be {\mathcal T}= \int dJ (h_{in}'-h_{out}') \lb{dJA-ch} \ee
with $h'=h+\dot{\phi},$ which implies an instantaneous equality between energy 
transfer rate and work done by the total pressure field $h'.$ Note that energy dissipation 
due to viscosity and other non-potential forces removes energy only from the rotational motions. 
See Appendix \ref{app:huggins} for a brief, self-contained derivation of Huggins' result. 

\begin{figure}
\center
\includegraphics[width=.52\textwidth]{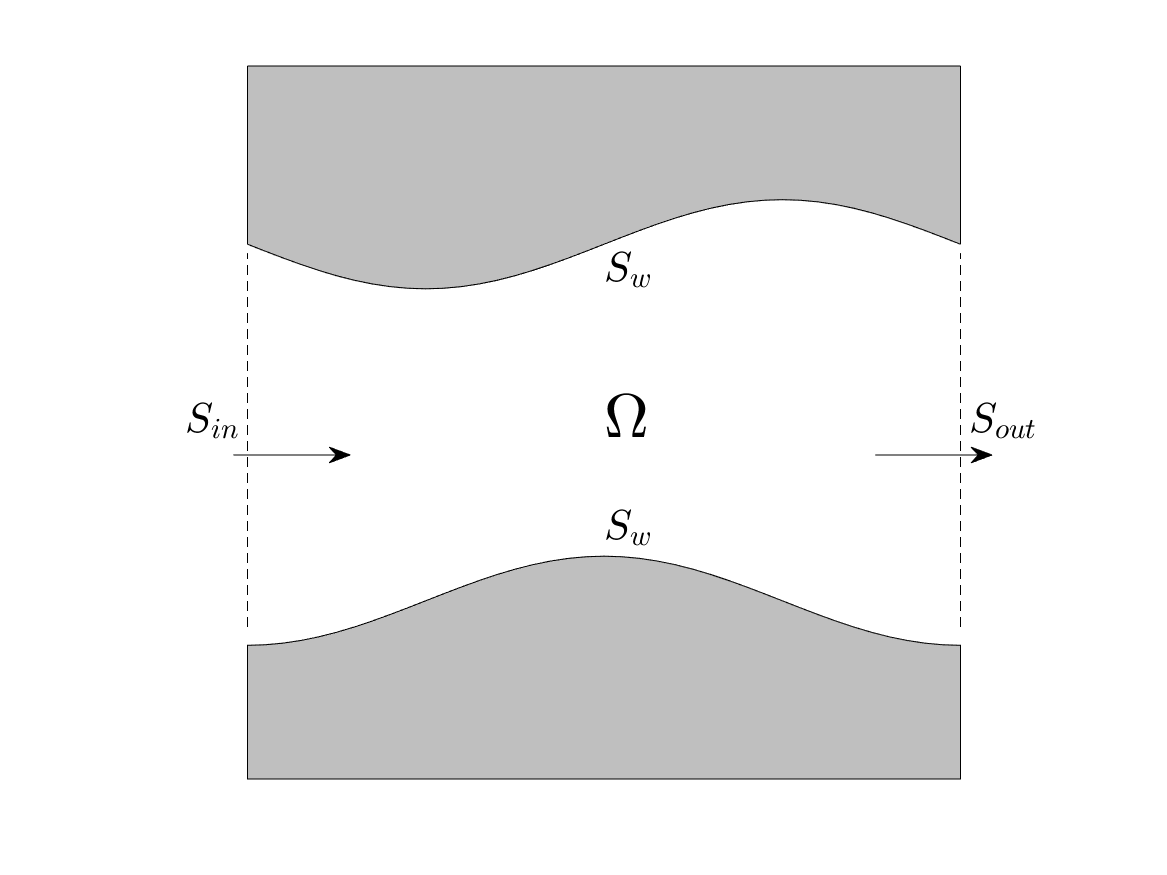}
\caption{Flow through a channel $\Omega$ with in-flow surface $S_{in},$ out-flow surface 
$S_{out},$  and channel walls $S_w$ } 
\lb{fig1} \end{figure}

The velocity decomposition introduced by Huggins is very natural for superfluids, where $\bu_\phi$ represents 
the ground-state superfluid velocity and $\bu_\omega$ the (incompressible) vortical excitations such 
as vortex rings. As we shall see in Section \ref{sec:pot-rot}, Lighthill  \cite{lighthill1963introduction} 
introduced the identical decomposition in his discussion of flow around a solid body, using quite 
different arguments. Furthermore, Lighthill recognized that there would be a creation of vorticity 
at solid walls with a normal flux $\bsigma$ related exactly to Huggins' vorticity flux tensor, as  
\be \bsigma= \hat{\bn}\bdot \bSigma =\hat{\bn}\btimes\nu(\grad\btimes\bomega) 
+\hat{\bn}\btimes\bg, \lb{lyman} \ee
which is now called the {\it Lighthill vorticity source}. In fact, Lighthill wrote his source explicitly 
for a flat wall only, and the general formula above for a curved wall was first proposed by Lyman 
\cite{lyman1990vorticity}. There is more than one possible generalization of Lighthill's flat-wall 
expression to curved walls (e.g. see \cite{panton1984incompressible,wu1993interactions,
wu1996vorticity,wu1998boundary}), but Lyman's proposal is uniquely the one that corresponds 
to creation of circulation at the boundary (\cite{eyink2008turbulent}, Appendix A). 
Lighthill  \cite{lighthill1963introduction} further realized that vorticity generation at the wall could be 
related to tangential pressure gradients, as can be seen by substituting the equation of motion
\eqref{NS-ext} into \eqref{lyman}, to obtain 
\be \bsigma = -\hat{\bn}\times(\grad h+\partial_t\bu). \lb{morton} \ee 
Here we have included the term $\partial_t\bu,$ which is non-zero if the wall is accelerating tangential
to itself, that was first introduced by Morton \cite{morton1984generation}, who also emphasized 
the inviscid nature of vorticity generation according to formula \eqref{morton}. 
In common with Josephson and Anderson, Lighthill \cite{lighthill1963introduction} thus realized 
also that pressure gradients and transverse vorticity fluxes are inextricably linked. 

\section{The Classical Josephson-Anderson Relation for a Finite Moving Body} 

In this section we present the derivation of the new Josephson-Anderson relation for flow around a body,
which requires extended mathematical discussion.  However, a reader who is most interested in concrete 
applications can skip directly to section \ref{sec:sphere} where flow around a sphere is considered 
in detail as an illustrative example. 

\subsection{Set-Up of the Problem}\lb{sec:setup}

We consider the flow around a finite solid body $B$ with smooth boundary $\partial B,$ held 
at rest, in an incompressible fluid that is filling the region $\Omega={\mathbb R}^3\backslash B$ 
and moving at constant velocity ${\it \bV}={\rm V}\hat{\bx}$ upstream of the body and at far distances
from it. See Figure \ref{fig2}. By Galilean invariance,  we can equivalently consider the body to be in translational 
motion with velocity $-{\rm V}\hat{\bx}$ through a fluid at rest and that point of view is sometimes more convenient. 
In this flow set-up we shall consider the solution $\bu$ of the viscous incompressible Navier-Stokes equation 
\be \partial_t\bu +\grad\bdot\left(\bu\bu+p\bI -2\nu\bS\right)=\bzed, 
\quad \grad\bdot\bu=0 \lb{NS-mom} \ee 
with the boundary conditions 
\be \left.\bu\right|_{\partial B}=\bzed, \quad \bu\underset{|\bx|\to\infty}{\sim} \bV \lb{NS-bc}. \ee
Note that in Eq.\eqref{NS-mom} we have written the Navier-Stokes equation as a local conservation law 
for linear momentum, with the total stress tensor (in dyadic notation) 
\be {\bf T}= \bu\bu+p\bI -2\nu\bS \lb{T-def} \ee
where $\bS=\frac{1}{2}[(\grad\bu)+(\grad\bu)^\top]$ is the strain-rate tensor. We shall 
assume here that the Navier-Stokes solutions are smooth for all times. 

For comparison, we shall also consider the potential-flow solution of the incompressible Euler 
equation 
\be \partial_t\bu_\phi +\grad\bdot\left(\bu_\phi\bu_\phi+p_\phi\bI\right)=\bzed, 
\quad \grad\bdot\bu_\phi=0 \lb{Euler} \ee 
with $\bu_\phi=\grad\phi$ given by a velocity potential $\phi$ which satisfies the boundary conditions
\be \left.\frac{\partial\phi}{\partial n}\right|_{\partial B}=0, \quad \phi\underset{|\bx|\to\infty}{\sim} {\rm V}x 
\lb{phi-bc} \ee 
Standard theory of potential flow implies that there is a unique, smooth solution with potential $\phi$
satisfying the Laplace equation $\triangle\phi=0$ for b.c. \eqref{phi-bc} and with kinematic pressure given 
by the Bernoulli equation
\be \partial_t\phi +\frac{1}{2} |\bu_\phi|^2+p_\phi = 0. \lb{bernoulli} \ee 
Here the pressure $p_\phi$ has been assumed for convenience to equal the constant value 
$-\frac{1}{2}{\rm V}^2$ at infinity.  This potential flow is, of course, the subject of the famous 
{\it D'Alembert paradox} \cite{dalembert1749theoria,grimberg2008genesis}
according to which the force exerted by the fluid on the body  
\be {\bf F}_\phi=-\int_{\partial B} P_\phi\, \hat{\bn} \, dA \lb{phi-drag} \ee 
has vanishing drag, or $\bV\bdot{\bf F}_\phi=0.$ Note in Eq.\eqref{phi-drag} and hereafter 
that $\hat{\bn}$ denotes the normal at the surface $\partial B$ pointing from the solid body 
into the fluid. 

\begin{figure}
\center
\includegraphics[width=.5\textwidth]{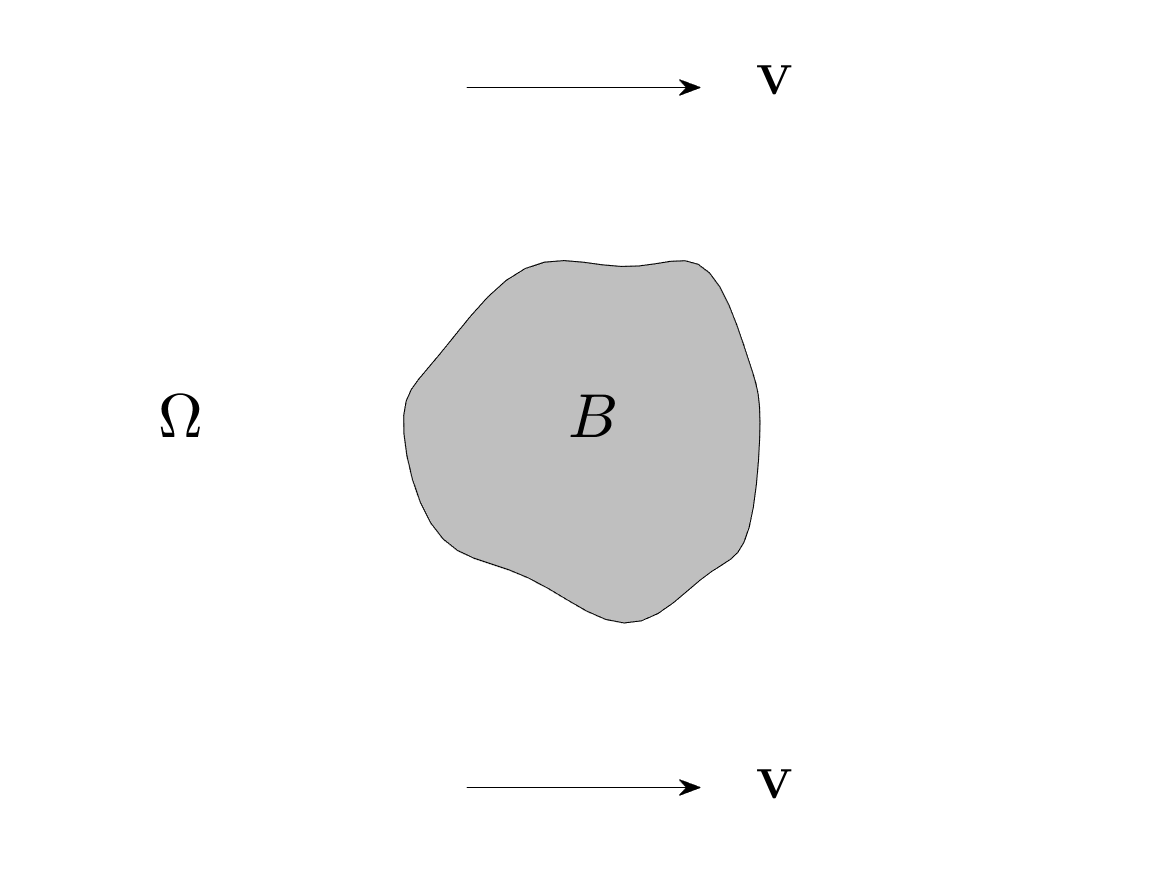}
\caption{Flow around a finite body $B$ in an unbounded region $\Omega$
filled with fluid moving at a velocity ${\bf V}$ at far distances.} 
\lb{fig2} \end{figure}

\subsection{Potential/Vortical Representation of Navier-Stokes}\lb{sec:pot-rot}  

Following the approach of Huggins \cite{huggins1970energy} to derive the 
Josephson-Anderson relation,  we introduce the corresponding 
rotational contributions by the definitions
\be \bu_\omega:=\bu-\bu_\phi,\quad p_\omega:=p-p_\phi. \lb{rot-def} 
\ee
Although the Navier-Stokes velocity field is thereby decomposed as 
$\bu=\grad\phi+\grad\btimes\bpsi_\omega$ into gradient and curl parts,
this is not the familiar Helmholtz decomposition. Recall that the Helmholtz decomposition 
for a piecewise smooth vector field $\bv$ which is zero inside $B$ and smooth in the 
external flow domain $\Omega$ takes the form: 
\begin{eqnarray} 
&& \bv(\bx)=-\grad\left(\int_\Omega \frac{(\grad\bdot\bv)(\bx')}{4\pi  |\bx-\bx'|}dV'
+\int_{\partial B} \frac{\hat{\bn}\bdot \bv(\bx')}{4\pi  |\bx-\bx'|}dA'\right)  \cr 
&& + \grad\btimes\left(  \int_\Omega \frac{(\grad\btimes\bv)(\bx')}{4\pi  |\bx-\bx'|}dV'
+\int_{\partial B} \frac{\hat{\bn}\btimes\bv(\bx')}{4\pi  |\bx-\bx'|}dA'
\right) \cr
&& 
\lb{helmholtz} \end{eqnarray}
See \cite{kustepeli2016helmholtz} for an engaging discussion of the history of the 
Helmholtz decomposition and a useful survey of the mathematical literature, and see also 
\cite{vladimirov1978vortical} for application to Navier-Stokes solutions outside a solid body. 
It is important to emphasize that the Helmholtz decomposition cannot be used to 
obtain uniquely the solenoidal velocity field $\bu$ corresponding to a given vorticity field $\bomega$
in $\Omega$ without specifying appropriate boundary conditions at $\partial B$ and,
furthermore, that arbitrary boundary conditions cannot be imposed.  Applied to a 
Navier-Stokes solution $\bu$ with b.c. \eqref{NS-bc}, the Helmholtz decomposition yields 
\be \bu(\bx,t) = \grad\btimes\left(  \int_\Omega \frac{\bomega(\bx',t)}{4\pi  |\bx-\bx'|}dV'  \right) 
\lb{u-helm} \ee 
As already noted by Lighthill, ``there is  only a restricted class of vorticity distributions that correspond 
to real flows satisfying also the no-slip condition'' \cite{lighthill1963introduction}. 

The velocity field \eqref{u-helm} is not, however, the only one that yields the vorticity distribution 
$\bomega=\grad\btimes\bu.$
As also noted by Lighthill, ``for {\it any} given solenoidal distribution of vorticity $\bomega$
outside the body (whose motion is again prescribed), one and only one solenoidal velocity
field exists, tending to zero at infinity and with zero normal velocity relative to the surface''
\cite{lighthill1963introduction}. In fact, this unique velocity field mentioned by Lighthill for the 
Navier-Stokes solution $\bu$ is exactly the field $\bu_\omega$ defined in \eqref{rot-def}, which by 
its definition satisfies the boundary conditions
\begin{eqnarray}  
\left.\hat{\bn}\bdot\bu_\omega\right|_{\partial B}=0, && \hspace{-20pt} 
\left.\hat{\bn}\btimes\bu_\omega\right|_{\partial B} = -\left.\hat{\bn}\btimes\bu_\phi\right|_{\partial B}\cr
&&\cr 
\bu_\omega &\underset{|\bx|\to\infty}{\sim} & \bzed, 
\lb{uom-bc} \end{eqnarray} 
as a consequence of \eqref{NS-bc},\eqref{phi-bc}. The Helmholtz decomposition for this 
velocity field $\bu_\omega$ yields the representation
\be \bu_\omega(\bx,t) = \grad\btimes\left(  \int_\Omega \frac{\bomega(\bx',t)}{4\pi  |\bx-\bx'|}dV' 
+\int_{\partial B} \frac{\hat{\bn}\btimes\bu_\omega(\bx',t)}{4\pi  |\bx-\bx'|}dA'
 \right) \lb{uom-helm} \ee 
and likewise the Euler solution $\bu_\phi$ is given by 
\be \bu_\phi(\bx,t)=
\grad\btimes\left(\int_{\partial B} \frac{\hat{\bn}\btimes\bu_\phi(\bx',t)}{4\pi  |\bx-\bx'|}dA'
\right),  \lb{uphi-helm} \ee 
which is an alternative representation of the potential flow as $\bu_\phi=\grad\btimes\bpsi_\phi$ 
in terms of a vector potential. 

 The quantities $\hat{\bn}\btimes\bu_\omega,$ $\hat{\bn}\btimes\bu_\phi$ in the integral 
 representations \eqref{uom-helm},\eqref{uphi-helm} have a simple physical interpretation as 
 singular vorticity sheets on the surface $\partial B,$ equal and opposite to each other.  
 This observation allows us to identify the fields $\bu_\om,$ $\bu_\phi$ introduced by Huggins 
 \cite{huggins1970energy} with corresponding fields that appear in Lagrangian vortex 
 methods for solving the Navier-Stokes equation, which were pioneered by Payne 
 \cite{payne1956numerical,payne1958calculations}  and advocated by Lighthill  \cite{lighthill1963introduction}. 
These methods have since been substantially developed with many alternative schemes 
proposed. See \cite{ploumhans2002vortex} for a clear discussion of the application to flow around 
a solid body and see \cite{branlard2017different} for a recent general review. In the version
explained by Lighthill  \cite{lighthill1963introduction}, given a vorticity distribution $\bomega$ 
in $\Omega$ at each instant (obtained by advecting, stretching and diffusing 
the prior distribution), one uses the Biot-Savart formula to construct the unique solenoidal 
velocity field which is vanishing at infinity and satisfying no-flow-through conditions at $\partial B,$
which is the field $\bu_\omega.$ However, this velocity field does not satisfy the stick b.c at $\partial B$
on its tangential components and, also, it is not a constant velocity $\bV$ at infinity in the body frame. 
To remedy these defects, one must add a tangential vortex sheet $\hat{\bn}\btimes\bu_\phi$ at the body 
surface $\partial B,$ so that the resultant velocity field $\bu=\bu_\omega+\bu_\phi$ satisfies both conditions. 
More physically, $\hat{\bn}\btimes\bu_\phi$ is considered as the newly generated vorticity at the surface
which, as pointed out by Lighthill, corresponds to vorticity oriented along equipotential lines of the Euler flow. 

We take here, however, a very different point of view. Since the Euler flow $\bu_\phi$ satisfying 
the b.c. \eqref{phi-bc} is unique and easily computed, either analytically or numerically, 
we shall instead use the definitions \eqref{rot-def} to introduce a new formulation of 
incompressible Navier-Stokes which we call the {\it potential/vortical formulation}. 
This can be easily obtained by taking the difference of the Navier-Stokes and Euler 
equations to obtain an equation of motion for $\bu_\omega,$ of the form 
\be \partial_t\bu_\omega +\grad\bdot\left(\bu_\omega\bu_\omega+\bu_\phi\bu_\omega
+\bu_\omega\bu_\phi+p_\omega\bI -2\nu\bS\right)=\bzed.  \lb{NS-omega-mom} \ee 
This must be solved with the b.c. 
\be \left.\bu_\omega\right|_{\partial B}=-\left.\bu_\phi\right|_{\partial B}, 
\quad \bu_\omega\underset{|\bx|\to\infty}{\sim} \bzed. \lb{u-om-bc} \ee 
and also the pressure $p_\omega$ chosen to enforce the incompressibility condition 
$\grad\bdot\bu_\omega=0.$ From the obtained $\bu_\omega$, the Navier-Stokes solution can then be 
reconstructed as 
\be \bu=\bu_\phi+\bu_\omega. \lb{u-reconstr} \ee
The equation \eqref{NS-omega-mom} can be regarded as expressing the local conservation of 
the integral $\bP_\omega=\rho\int_\Omega \bu_\omega\,dV,$ which we shall call the 
{\it vortex momentum}.  Of course, other equivalent forms of equation \eqref{NS-omega-mom} 
can be derived. 
For example, substituting $\partial_t\bu_\phi=\grad\dot{\phi}$ and using vector calculus identities 
yields 
\begin{eqnarray}
\partial_t\bu_\omega &=&\bu\btimes\bomega-\nu\grad\btimes\bomega +
\grad\left(p+\frac{1}{2}|\bu|^2+\dot{\phi} \right). \cr 
&=& \bu\btimes\bomega-\nu\grad\btimes\bomega +
\grad\left(p_\omega+\frac{1}{2}|\bu_\omega|^2+\bu_\omega\bdot\bu_\phi\right), \cr
&& 
\lb{NS-omega-mom2} \end{eqnarray} 
where in the second line the Bernoulli equation \eqref{bernoulli} was invoked. This 
version of the potential/vortical formulation is more physically intuitive in terms of 
vortex dynamics and shall be our main tool in this work. However, this version contains 
expressions such as $\bu\btimes\bomega$ which are hard to give rigorous meaning 
when $Re\to\infty,$ and thus the conservation form \eqref{NS-omega-mom} is preferred 
in considering the infinite Reynolds-number limit. 

As should be clear from the review in section \ref{sec:review}, the detailed Josephson-Anderson
relation involves as well the conservation of vorticity and kinetic energy. The equation expressing 
local conservation of vorticity can be obtained in the potential/vortical formulation by taking the 
curl of \eqref{NS-omega-mom2}. It has the same form as \eqref{vort-cons} with the 
vorticity-flux tensor $\bSigma$ as given in \eqref{Sigma} with $\bg\equiv \bzed.$ This is, 
of course, just the usual Helmholtz equation. 

The equation for the local conservation of the kinetic energy of the rotational flow can be obtained 
by dotting $\bu_\omega$ into \eqref{NS-omega-mom2}, which yields 
 \begin{eqnarray}
&& \partial_t\left(\frac{1}{2}|\bu_\omega|^2\right) +
\grad\bdot\left[\left(p_\omega+\frac{1}{2}|\bu_\omega|^2 +\bu_\omega\bdot\bu_\phi\right)\bu_\omega-
\nu\bu\btimes\bomega\right] \cr
&& \qquad\qquad\qquad = -\bu_\phi\bdot(\bu\btimes\bomega-\nu\grad\btimes\bomega)-\nu|\bomega|^2 
\lb{Eom-loc} \end{eqnarray} 
Note that the $\bu_\phi\bdot(\nu\grad\btimes\bomega)$ term is needed on the righthand side 
of \eqref{Eom-loc} so that $\nu\bu\btimes\bomega$ will appear in the square bracket on the lefthand side. 
Otherwise, the term inside the square bracket would be $\nu\bu_\omega\btimes\bomega,$ whose normal 
component does not vanish on the surface $\partial B$ of the object. Since the expression in the square bracket 
is a spatial energy flux, it should imply a vanishing flux through the surface. A corresponding equation can be 
obtained also for the {\it interaction energy} of potential and vortical flow, by dotting $\bu_\phi$ into 
\eqref{NS-omega-mom2}: 
\begin{eqnarray}
&& \partial_t\left(\bu_\phi\bdot \bu_\omega \right) +
\grad\bdot\left[\left(p_\omega+\frac{1}{2}|\bu_\omega|^2 +\bu_\omega\bdot\bu_\phi\right)\bu_\phi \right. \cr 
&& +\left.\left(p_\phi+\frac{1}{2}|\bu_\phi|^2\right) \bu_\omega\right] 
= +\bu_\phi\bdot(\bu\btimes\bomega-\nu\grad\btimes\bomega) \cr 
&& 
\lb{Eint-loc} \end{eqnarray} 
The equal and opposite terms on the righthand sides of Eqs.\eqref{Eom-loc},\eqref{Eint-loc} 
clearly represent energy transfer between rotational and potential flow. 
Note that the triple product $\bu_\phi\bdot(\bu\btimes\bomega)$ can be rewritten as 
$\bu_\phi\bdot(\bu_\omega\btimes\bomega)$ using $\bu=\bu_\omega+\bu_\phi,$ so that 
only self-advection of vorticity by the rotational motions themselves contributes to transfer. 

Because of the cancellation of these two terms, the sum of the rotational and interaction energies
satisfies the equation 
\begin{eqnarray}
&& \partial_t\left(\frac{1}{2}|\bu_\omega|^2+ \bu_\phi\bdot \bu_\omega\right) \cr
&& 
+ \grad\bdot\bigg[\left(p_\omega+\frac{1}{2}|\bu_\omega|^2 +\bu_\omega\bdot\bu_\phi\right)\bu 
+\left(p_\phi+\frac{1}{2}|\bu_\phi|^2\right) \bu_\omega \cr 
&& \qquad \qquad \qquad \qquad \qquad -\nu\bu\btimes\bomega \bigg] \ = \ -\nu|\bomega|^2.  \lb{Etot-loc} 
\end{eqnarray} 
Importantly, this combination of energies is conserved for $\nu\to 0,$ as long as solutions 
remain smooth in the limit. For this reason, we shall refer to the combined quantity $E(t)=\rho\int_\Omega \left[ 
\frac{1}{2}|\bu_\omega|^2+\bu_\omega\bdot\bu_\phi\right]$ as the {\it total kinetic energy}. 
Of course, this is not literally true, because the kinetic energy of the potential flow 
$\frac{1}{2}\rho|\bu_\phi|^2$ is missing. However, the latter energy is separately conserved by the 
Euler equation \eqref{Euler} for $\bu_\phi,$ so that this contribution can be neglected when 
considering the total energy balance involving the rotational motions. 

The detailed Josephson-Anderson relation for confined channel flow that was considered in section 
\ref{sec:review} involves the global balances of kinetic energy, not the local ones. Before we can derive 
the analogue of that relation for our problem, we must consider vorticity generation at the body surface
and the precise far-field asymptotics of the rotational velocity field. 

\subsection{Generation of Vorticity at the Boundary}\lb{sec:generate} 

In order to make certain that our problem is physically meaningful, we must consider an issue
neglected until now, namely, the acceleration of the body from rest to constant velocity. 
Here it is more natural to consider the body as moving and the fluid as at rest at infinity. 
This relative motion can be accomplished by a translational acceleration protocol ${\bf a}(t)$ 
which over some time interval $0\leq t\leq T$ takes the body from zero velocity to velocity ${\bf V},$
including the possibility of an impulsive acceleration ${\bf a}(t)=\bV\delta(t)$ with $T=0.$
The Lighthill vorticity source for this situation is of the form \cite{morton1984generation} 
\be \bsigma=-\hat{\bn}\btimes ({\bf a}(t) + \grad p)  \ee 
where it is assumed that ${\bf a}(t)\equiv \bzed$ for $t>T.$ It follows that vorticity 
is created only tangential to the body surface and, after the time $T,$ is generated along 
the surface isobars or pressure isolines \cite{lighthill1963introduction}. Here it is appropriate 
to note that, because of the stick boundary conditions on the velocity, $\hat{\bn}\bdot\bomega=0$ 
everywhere on the surface of any non-rotating body $B.$ Thus, vortex lines on solid 
surfaces in classical Newtonian fluids lie in general {\it parallel} to the surface $\partial B.$
Vortex lines can only terminate on a solid (non-rotating) wall at some exceptional points where 
$\bomega=\bzed,$ which are possible points of boundary-layer separation \cite{lighthill1963introduction}.
This is an important difference from superfluids, where quantized vortex lines can often terminate 
at a solid surface and, when they do so, intersect it almost normally so as to satisfy the no-flow-through 
condition \cite{schwarz1985three}. Vortex half-rings can in fact be observed standing at the surface of 
a body moving through a superfluid (e.g. see Fig.2 in \cite{winiecki2000motion}). On the contrary, vorticity
is generated principally parallel to the surface in the classical case, as closed vortex loops encircling the body. 

If the fluid is initially at rest or, more generally, has zero net vorticity, then this condition is 
preserved in time:
\be  \int_\Omega \bomega \, dV = \bzed.  \ee
This result is known as F\"oppl's theorem \cite{foppl1897geometrie}, but the standard 
proof (e.g. see \cite{wu2007vorticity}, p.74) assumes some sufficient decay of the vorticity at infinity,  
which is {\it a priori} unknown in our problem. It is therefore important and illuminating to 
give a direct proof based upon Lighthill's theory, by showing that 
\be \frac{d}{dt}\int_\Omega \bomega \, dV = -\int_{\partial B} \hat{\bn}\btimes ({\bf a}(t) + \grad p) dA 
=\bzed. \ee 
The first term involving the spatially uniform acceleration ${\bf a}(t)$ easily vanishes due to 
the elementary result \cite{wu2007vorticity}
\be \int_{\partial B} \hat{\bn}\,dA=\bzed. \ee
The term involving the pressure gradient can be rewritten as 
\be \int_{\partial B} \hat{\bn}\btimes \grad p\, dA =  \int_{\partial B} \hat{\bt}\, |\grad p|\, dA \ee 
where
$$ \hat{\bt}= \hat{\bn}\btimes \frac{\grad p}{|\grad p|} =\frac{d\bx}{ds}(p,s) $$
is the tangent vector to surface isobars and $\bx(p,s)$ is a smooth parameterization of the 
isobar with pressure value $p,$ in terms of arclength $s$ along the isobar.  Here we 
appeal to the Sard theorem of differential topology which implies for a smooth pressure 
field $p(\bx,t)$ that, for a.e. $p,$ the connected components of the surface isobar with that $p$-value 
are simple closed smooth curves on $\partial B$ \cite{hirsch2012differential}.  Then
using the standard result
\be  \int_{\partial B} \delta(p(\bx,t)-p) dA = \int_{\{\bx:\, p(\bx,t)=p\}} \frac{ds}{|\grad p(\bx(p,s),t)|} \ee 
known in mathematics as the coarea formula \cite{federer2014geometric},  
it follows that 
$$ \int_{\partial B} \hat{\bn}\btimes \grad p\, dA = 
     \int_{\partial B} \hat{\bt}(p,s) \, dp \,ds 
     =\int dp \int ds \frac{d\bx}{ds} = \bzed $$
since the isobars are closed curves for a.e. $p$-value. 

\subsection{Far-Field Velocities and Vortex Impulse}\lb{sec:far} 

With the previous results in hand, we now develop asymptotic formulas for the velocities 
$\bu_\omega(\bx,t)$ and $\bu_\phi(\bx,t)$ in the far-field, or for large $r=|\bx|.$ 

We begin with the rotational velocity field $\bu_\omega.$ Here we follow closely an 
argument of Cantwell for a different problem of forced jets \cite{cantwell1986viscous},  
by considering the vector potential that appears in the Helmholtz formula \eqref{uom-helm}
\be \bpsi_\omega(\bx,t)=\int_\Omega \frac{\bomega(\bx',t)}{4\pi  |\bx-\bx'|}dV' 
+\int_{\partial B} \frac{\hat{\bn}\btimes\bu_\omega(\bx',t)}{4\pi  |\bx-\bx'|}dA'  \ee 
so that $\bu_\omega=\grad\btimes\bpsi_\omega.$  
As in \cite{cantwell1986viscous}, we make a 
multipole expansion using $\frac{1}{|\bx-\bx'|}=\frac{1}{r}+\frac{\bx\bdot\bx'}{r^3}+\cdots$
obtaining 
\be \bpsi_\omega(\bx,t) = \frac{{\bf q}_\omega(t)}{4\pi r} + \frac{\bI_\omega(t)\btimes\bx}{4\pi r^3} +
O\left(\frac{1}{r^3}\right) \cdots  \lb{multipole}\ee 
where we have introduced the {\it total vorticity}  
\be {\bf q}_\omega(t) = \int_\Omega \bomega(\bx,t)\, dV
+\int_{\partial B} \hat{\bn}\btimes\bu_\omega(\bx,t)\, dA, \lb{q-def} \ee 
including the contribution from the surface vortex sheet, and 
also the corresponding {\it vortex impulse} \cite{huggins1971dynamical, vladimirov1978vortical}  
\be \bI_\omega(t) = \frac{1}{2}\left[\int_\Omega \bx\btimes \bomega(\bx,t)\, dV
+\int_{\partial B} \bx\btimes(\hat{\bn}\btimes\bu_\omega(\bx,t))\, dA \right] \lb{I-def} \ee 

We first note that ${\bf q}_\omega(t) \equiv\bzed$ so that the ``monopole'' term 
in Eq.\eqref{multipole} is zero. The volume integral in the definition \eqref{q-def} 
of ${\bf q}_\omega$ vanishes by the results of the previous section \ref{sec:generate}. 
Furthermore, by the boundary condition for $\bu_\omega$ on $\partial B,$ 
\be \int_{\partial B} \hat{\bn}\btimes\bu_\omega(\bx,t) \, dA=
- \int_{\partial B} \hat{\bn}\btimes\grad\phi(\bx,t)\, dA=\bzed \ee
by precisely the same argument as in section \ref{sec:generate}.  

We thus obtain by a curl of \eqref{multipole} the final result 
\be
\bu_\omega(\bx,t) \underset{|\bx|\to\infty}{\sim} \frac{-\bI_\omega(t)r^2+3(\bI_\omega(t)\bdot\bx)\bx}{4\pi r^5} 
\lb{u-om-far1} \ee 
which is a dipole field. Quite intuitively, the vortical wake behind the body appears at very large distances 
like a vortex ring with impulse $\bI_\omega(t)$. The formula \eqref{u-om-far1} can be simply rewritten in spherical coordinates
for polar angle $\theta$ measured from the positive $x$-axis, as:  
\be 
\bu_\omega(\bx,t) 
\underset{|\bx|\to\infty}{\sim} \frac{{\rm I}_\omega}{4\pi r^3}\hat{\bx} - \frac{3{\rm I}_\omega\cos\theta}{4\pi r^3}\hat{\br} 
\lb{u-om-far} \ee 
with $\bI_\omega=-{\rm I}_\omega\hat{\bx}.$ The sign here can be simply understood because the vortical wake 
behind the body must reduce the velocity of the potential flow. Alternatively, in the fluid rest frame, 
the vortical impulse must be in the direction of motion of the body. 

Asymptotics of the potential flow velocity can be similarly obtained from $\bu_\phi=\grad\btimes\bpsi_\phi$
where
\be \bpsi_\phi(\bx,t)=\int_{\partial B} \frac{\hat{\bn}\btimes\bu_\phi(\bx',t)}{4\pi |\bx-\bx'|}dA' \ee  
This velocity to leading order is the constant ${\rm V}\hat{\bx}$ plus a dipole term 
similar to Eq.\eqref{u-om-far1}, but involving the vortex impulse ${\bf I}_\phi(t)$
associated to the surface discontinuity. However, in what follows 
we shall only need the leading term, so that 
\be \bu_\phi(\bx,t) \underset{|\bx|\to\infty}{\sim} {\rm V} \hat{\bx} + O\left(\frac{1}{r^3}\right) \lb{u-phi-far} \ee 
or equivalently in terms of the scalar potential 
\be \phi(\bx,t) \underset{|\bx|\to\infty}{\sim} {\rm V} r\cos\theta + O\left(\frac{1}{r^2}\right).  \lb{phi-far} \ee 

\subsection{Global Momentum and Energy Integrals}\lb{sec:integrals} 

Using the asymptotic formulas of the preceding section \lb{sec:far} we can 
now study the global integrals of momentum and kinetic energy for our flow. 

The total vortex momentum is defined by
\be \bP_\omega(t) = \rho \int \bu_\omega(\bx,t) \, dV. \ee 
With the dipole asymptotics \eqref{u-om-far1} for the integrand, this integral is convergent 
but only conditionally so. Note that the vortex momentum is not equal to density times 
vortex impulse. In fact, it is well-known that 
\be \bP_\omega(t) = \frac{2}{3}\rho \bI_\omega(t). \lb{cantwell} \ee
This can be shown by an argument of Cantwell \cite{cantwell1986viscous}, using the representation
$\bu_\omega=\grad\btimes\bpsi_\omega$ and the identity
$\int_{S_R} \frac{\bx}{4\pi|\bx-\bx'|}d\Omega=\frac{1}{3}\frac{\bx'}{R}$ for 
a sphere $S_R$ of radius $R>r'.$ In the Appendix \ref{app:vortmom} we give 
another proof. 

Total kinetic energy of the vortical wake is clearly well-defined by the integral 
\be E_\omega(t)=\frac{1}{2}\rho\int_\Omega |\bu_\omega(\bx,t)|^2\, dV,\ee 
because the square of the dipole field decays $\sim 1/r^6.$ However, the 
total interaction energy 
\be E_{int}(t) = \rho\int_\Omega \bu_\omega(\bx,t)\bdot\bu_\phi(\bx,t)\, dV \ee 
is again at most conditionally convergent and, if convergent, might be expected to vanish! 
As we recall, the Kelvin minimum energy theorem is exactly the statement that the
potential and vortical velocity fields are orthogonal and, in his discussion of a 
moving body, Lighthill appealed to this result (see \cite{lighthill1963introduction}, p.56).
However, he worked in the rest frame of the fluid, where the Kelvin minimum energy theorem 
indeed holds, but we work in the rest frame of the body where it does not.  

We find  instead that 
\be E_{int}(t)=\bP_\omega(t)\bdot \bV. \ee 
The proof is simple: using $\bu_\omega\bdot\bu_\phi=\grad\bdot(\phi\bu_\omega),$ we obtain 
\begin{eqnarray} 
E_{int}(t) &=& \lim_{R\to\infty} \rho \int_{B_R\backslash B}  \grad\cdot(\phi\bu_\omega) \, dV \cr 
&=& \lim_{R\to\infty} \rho \int_{S_R}  \phi(\hat{\br}\bdot \bu_\omega) \, dA
\end{eqnarray} 
where $B_R$ is the ball of radius $R$ centered at the origin so that $\partial B_R=S_R.$
Note that the contribution from the body surface $\partial B$ vanishes because $\hat{\bn}\bdot\bu_\omega=0$
there. Using the asymptotics Eq.\eqref{u-om-far},\eqref{phi-far} for $\bu_\om$ and $\phi,$
gives the integral over solid angle $d\Omega=\sin\theta\,d\theta\,d\varphi$
\begin{eqnarray}
E_{int}(t) &=& \lim_{R\to\infty} \rho \int_{S_R}  {\rm V}R\cos\theta 
\left(-\frac{{\rm I}_\omega\cos\theta}{2\pi R^3}\right) R^2 \, d\Omega\cr
&=& -\frac{2}{3}\rho {\rm I}_\omega{\rm V} \ = \ \frac{2}{3}\rho \bI_\omega\bdot \bV \ = \ \bP_\omega\bdot\bV.  
\lb{Eint} \end{eqnarray} 

We conclude finally that the total energy in the body frame is given by 
\be E(t) = E_\omega(t) + \bP_\omega(t)\bdot\bV, \ee 
a result that should be expected by Galilean invariance.  This conclusion will appear very 
familiar to superfluid physicists, since a similar result holds for a vortex ring moving in 
unbounded space \cite{schwarz1975onset,varoquaux2015anderson}. However, in that 
case $E_{int}=\rho \bI_\omega\bdot\bV$ in our notations \footnote{Note that  
\cite{schwarz1975onset,varoquaux2015anderson} and the superfluid community 
in general use the notation $\bP$ for the vortex impulse $\rho \bI$!}, 
i.e. it is vortex impulse which appears rather than vortex momentum. 

\subsection{Global Balances of Momentum and Energy}\lb{sec:gbal} 

We can now derive the balance equations for the global integrals in section \ref{sec:integrals} 
by integrating the local balance equations over $\Omega$.  Integrating Eq.\eqref{NS-omega-mom} 
gives the global momentum balance as  
\be \frac{d\bP_\omega}{dt} = {\bf F} \lb{dPdt} \ee
where 
\be {\bf F}= \int_{\partial B} (P_\omega\hat{\bn}+2\eta\bS\hat{\bn}) dA \ee 
is the total force applied to the fluid by the body. Note that all contributions from infinity 
are vanishing because the momentum-flux $\sim r^{-3}$ from the asymptotics 
\eqref{u-om-far},\eqref{u-phi-far}.  
This force can be rewritten also as 
\be {\bf F}=\int_{\partial B} (P_\omega\hat{\bn}+\eta\hat{\bn}\btimes\bomega) dA  \lb{F-vort} \ee 
since the viscous Newtonian stress vector at the wall $\btau_w=-2\eta\bS\hat{\bn}$ is related 
to the wall vorticity $\bomega_w$ by $\btau_w =\eta\bomega_w\btimes\hat{\bn}$ 
\cite{lighthill1963introduction,wu2007vorticity}.  An importance consequence of 
\eqref{dPdt} is that the total momentum of the vortical wake increases monotonically in time. 
In particular, there can be no global statistical steady-state for this flow. 

Integrating \eqref{Eint-loc} gives the global balance of the interaction energy as 
\be \frac{dE_{int}}{dt} = +\rho\int_\Omega \bu_\phi\bdot(\bu\btimes\bomega-\nu\grad\btimes\bomega) dV, 
\lb{Eint-bal} \ee 
where the asymptotics of the energy-flux $\sim r^{-3}$ implies no contribution from infinity and 
the no-flow-through condition implies no contribution from the body surface. 
On the other hand, directly differentiating expression \eqref{Eint} for interaction 
energy and using the global momentum balance \eqref{dPdt} gives \footnote{Notice that, 
until this point, none of our analysis requires that the velocity at infinity $\bV(t)$ must 
be constant in time, and so applies to more general cases of a body linearly accelerating 
through a fluid, but without body rotation. The Bernoulli equation \eqref{bernoulli} simply requires 
a new term $\dot{\bV}(t)\bdot\bx$ on the righthand side 
to obtain the Euler equation in the non-inertial body-frame. Body rotation, on the other hand,  
leads to new effects that are beyond the scope of the present paper.},
\be \frac{dE_{int}}{dt} = {\bf F}\bdot\bV. \lb{FV} \ee 
Although ${\bf F}$ must fluctuate in time, the dot-product above will generally be negative, as 
the force applied by the body opposes the fluid flow. 
Thus, drag appears as loss of energy of the potential flow, due to negative 
work by the body on the fluid. Note that $E_{int}(t)$ is monotonically decreasing, but 
the energy $E_{int}(t)+E_\phi$ does not decrease, of course, 
because there is an infinite reservoir of kinetic energy in the potential flow. 
The energy is transferred to the vortical 
flow, as can be seen by integrating Eq.\eqref{Eom-loc} in like fashion over $\Omega$ to give 
\be \frac{dE_\omega}{dt} = -\rho\int_\Omega \bu_\phi\bdot(\bu\btimes\bomega-\nu\grad\btimes\bomega) \, dV 
-\int_\Omega \eta|\omega|^2\, dV. \lb{Eom} \ee 
The energy transferred to the rotational fluid motions is ultimately dissipated by viscosity. 

The combination of \eqref{Eint-bal},\eqref{FV} yields the most fundamental result of our paper,
the classical version of the {\it detailed Josephson-Anderson relation} for flow past a solid body,  
written in various equivalent forms as 
\begin{eqnarray} 
-{\bf F}\bdot\bV &=& -\rho\int_\Omega \bu_\phi\bdot(\bu\btimes\bomega-\nu\grad\btimes\bomega) dV \cr 
&=&-\int dJ \int (\bu\btimes\bomega-\nu\grad\btimes\bomega)\bdot d\Bell \cr
&=& -\frac{1}{2} \int dJ \int \epsilon_{ijk} \Sigma_{ij} d\ell_k,  
\lb{dJA-bd} \end{eqnarray} 
where the second two expressions follow the notations in section \ref{sec:review}.  
The relation \eqref{dJA-bd} expresses an instantaneous balance between power injected by the drag
force ${\bf F}$ acting back on the fluid and the vorticity flux crossing the mass current along potential 
flow lines. This should be compared with the detailed Josephson-Anderson relation \eqref{dJA-ch}
derived by Huggins, which expresses an instantaneous equality between the rate of work done on 
the ideal potential flow by the total pressure $h'$ and the vorticity flux across the mass current, 
which thereby transfers that energy to vortical motions,  exactly as here. 

Although the momentum of the vortical motions is constantly increasing, Eq.\eqref{Eom} makes it 
plausible that the rotational flow energy should have a long-time steady-state limit,  sufficiently long after 
the initial acceleration of the body when the potential flow solution $\bu_\phi$ becomes time-independent. 
In fact, it is generally expected that the entire flow within any fixed distance of the body,
after a sufficiently long time that depends upon the distance, shall reach a steady-state whose character 
depends upon the Reynolds number, with a deterministic stationary flow at low Reynolds numbers, 
then periodic flow, and finally a chaotic flow with a long-time ergodic behavior at sufficiently high Reynolds 
number. Of course, there may be multiple distinct stable regimes, each attracting some domain of initial 
conditions. We shall refer to these hypothesized quasi-steady regimes in some vicinity of the body as the 
{\it local steady-states}.  At any finite time, however, one can observe at some far distance downstream 
a time-dependent flow with increasing momentum. 

Denoting the suitable time-average for such a local steady regime as $\langle\cdot\rangle$
and assuming that total vortical energy does in fact achieve a long-time mean value,
we can then average \eqref{Eom} and \eqref{dJA-bd} to obtain a steady-state version 
of the Josephson-Anderson relation as 
\begin{eqnarray} 
-\langle{\bf F}\rangle\bdot\bV &\doteq & -\rho\int_{\Omega'} \langle \bu_\phi\bdot(\bu\btimes\bomega-\nu\grad\btimes\bomega)\rangle \, dV \cr
&\doteq &  \eta \int_{\Omega'} \langle|\bomega|^2\rangle\, dV, \lb{meanJA} 
\end{eqnarray} 
where $\Omega'\subset \Omega$ has here been chosen large enough so that the space-integrals
over $\Omega'$ and $\Omega$ agree to any desired precision \footnote{It is important to note that 
all of the spatial integrals over $\Omega$ are absolutely convergent, because of the decay laws 
$\bu_\omega\btimes\bomega\sim r^{-7},$ $\grad\btimes\bomega\sim r^{-5},$ and $|\bomega|^2\sim r^{-8}$ 
asymptotically at large-$r$. Thus, the integrands both in the detailed Josephson-Anderson relation \eqref{dJA-bd} and in 
the global viscous dissipation in \eqref{Eom} are rather well-localized and the integrals can be calculated accurately 
in a possibly large, but finite-volume region of space.}, denoted by ``$\doteq$'', 
and then time-averages are taken over a long enough interval to obtain a steady-state within the 
region $\Omega'.$ The relation \eqref{meanJA} expresses equality in the mean of three 
distinct quantities: the power input by the drag force,  the energy transfer from potential to rotational 
motions by vortex motion, and the viscous energy dissipation. 
The third expression is clearly non-negative, expressing the time-irreversibility of the viscous
Navier-Stokes dynamics. It follows that the drag force must generally oppose the freestream 
velocity, just as expected. The middle expression in \eqref{meanJA} involves the vector quantity 
\be {\bf f}_v= \rho(\bu\btimes \bomega-\nu\grad\btimes \bomega)  \ee
which is sometimes called the {\it vortex force}, and, intuitively, drag is associated to the 
vortex force opposing the potential flow. An even more useful expression is 
\be
-\langle{\bf F}\rangle\bdot\bV=-\frac{1}{2} \int dJ \int \epsilon_{ijk} \langle\Sigma_{ij} \rangle d\ell_k 
\lb{meanJA2} \ee 
which represents mean drag in terms of vorticity flux crossing the potential mass current. 
In a local steady-state, mean vorticity flux further satisfies the relation dual to \eqref{gradh-Sigma} 
\be \langle\Sigma_{ij}\rangle =\epsilon_{ijk} \partial_k \langle h\rangle, \lb{Sigma-gradh} \ee 
implying also that $\partial_j\langle\Sigma_{ij}\rangle=0.$
Together, the two relations 
\eqref{meanJA2}, \eqref{Sigma-gradh} 
very strongly constrain the vortex dynamics and statistics that contribute to the mean drag. 

The detailed relation \eqref{dJA-bd} holds, of course, even before a local steady regime is 
achieved (but after the period of initial acceleration). One simple, general deduction can be 
made from this principle, by substituting the equation of motion \eqref{NS-omega-mom2}
into the energy transfer term to obtain \footnote{Note that the two integrands in the second line 
of \eqref{substitute} are separately only conditionally convergent, but their combination, 
which equals the integrand in the first line, is absolutely convergent.}
\begin{eqnarray}
-{\bf F}\bdot\bV &= & -\rho\int_\Omega \bu_\phi\bdot(\bu\btimes\bomega-\nu\grad\btimes\bomega) \, dV \cr 
&=& -\rho\int_\Omega \bu_\phi\bdot(\dot{\bu}_\omega+\grad h') \, dV 
\lb{substitute} \end{eqnarray}
with $h'=p+\frac{1}{2}|\bu|^2+\dot{\phi}$ as in \eqref{dJA-ch}.  
By applying the same arguments as those used to evaluate $\int_\Omega \bu_\phi\bdot\bu_\omega \, dV$ 
in section \ref{sec:integrals},  
\be -\rho\int_\Omega \bu_\phi\bdot\dot{\bu}_\omega \, dV = - \dot{\bP}_\omega(t)\bdot\bV = -{\bf F}\bdot \bV.  \ee 
The consequence is that 
\be \rho\int_\Omega \bu_\phi\bdot \grad h'\, dV = \int dJ (\Delta h') = 0 \lb{zero-head} \ee 
with the quantity 
\be \Delta h'= \int \grad h'\bdot d\Bell =h'(\ell=+\infty)-h'(\ell=-\infty) \ee
given as an integral along each streamline and representing the drop in the total pressure along 
this entire line. It seems certain that $\Delta h'\leq 0,$ since the presence of the body 
should cause a pressure drop and a reduced streamwise velocity in its wake. The conclusion 
from \eqref{zero-head} is that $\Delta h'=0$ along all of these lines, so that the total pressure 
recovers completely from whatever drop it experienced by the presence of the body. This 
result underlines the great difference from channel flow, where the detailed Josephson-Anderson 
relation \eqref{dJA-ch} derived by Huggins involves only the total press head in the channel. 

\section{Flow Past a Sphere}\lb{sec:sphere}  

To make the preceding discussion more concrete, we consider in this section the special
case of flow past a sphere. The rich phenomenology of this flow has been the subject of a 
recent review \cite{tiwari2020flow}, which classifies the flow into eight regimes as a function of 
increasing Reynolds number: (i) the axisymmetric wake regime, (ii) the planar symmetric wake 
regime with a counter-rotating pair of trailing streamwise vortices, (iii) the shedding regime with alternating hairpin vortices,
(iv) a regime with separating vortex-tubes due to Kelvin-Helmholtz instability of the separated 
boundary layer, (v) the subcritical regime where the point of instability moves upstream closer 
to the sphere, (vi) the critical regime of ``drag crisis'' with reattachment of the boundary layer 
and formation of a laminar separation bubble, (vii) the supercritical regime in which the bubble shrinks 
or disappears, and finally (viii) the transcritical regime, which is an apparently asymptotic state with constant 
mean drag. The detailed Josephson-Anderson relation \eqref{dJA-bd} holds in {\it all} of these 
regimes and reveals that, despite the very substantial differences in flow physics between the 
different regimes, the mechanism of drag in terms of vortex dynamics is intrinsically the same for all of them.

\subsection{Josephson-Anderson Relation for Sphere} 

\begin{figure}
\center
\includegraphics[width=.52\textwidth]{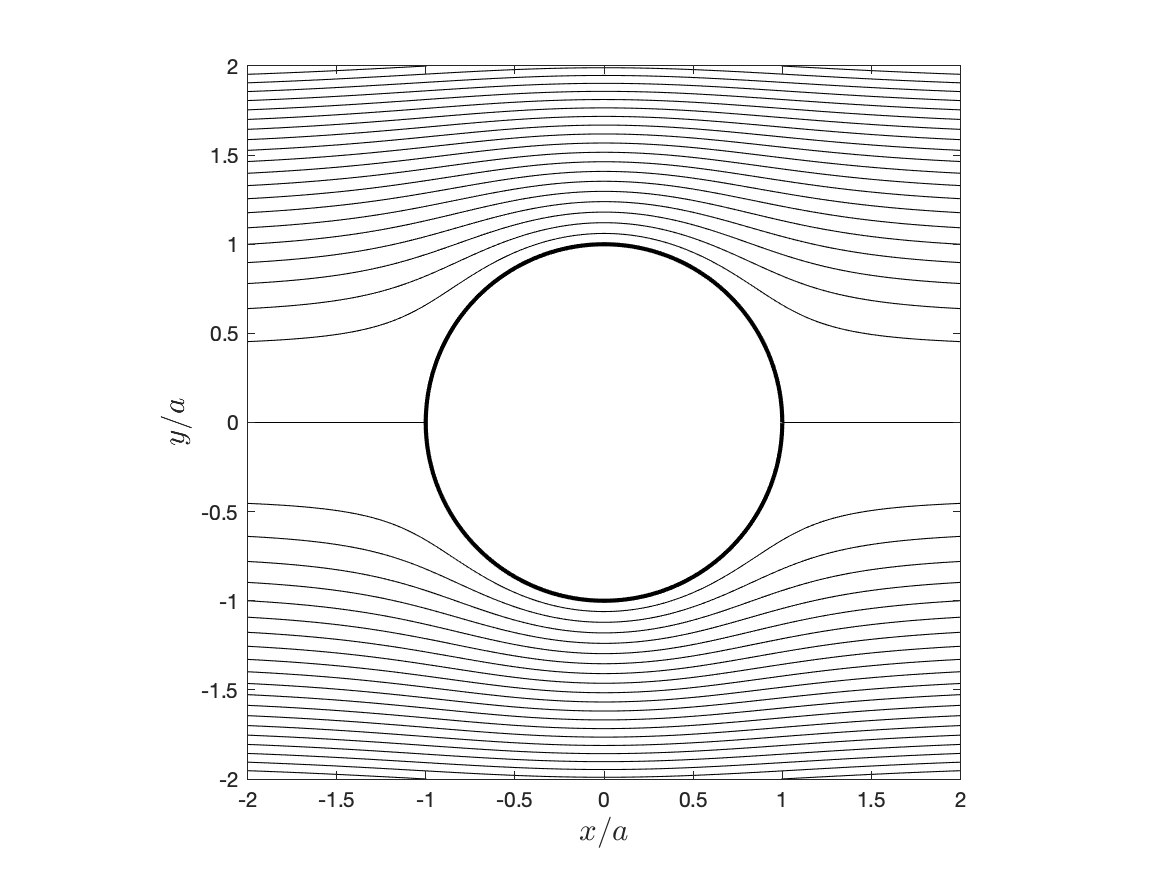}
\caption{Streamlines of the ideal flow around a sphere} 
\lb{fig3} \end{figure}

We give here the general relation \eqref{dJA-bd} a concrete form for flow around 
a sphere of radius $a$. The first important ingredient of that relation is the inviscid flow 
solution $\bu_\phi.$ This is well-known (\cite{batchelor2000introduction}, \S 6.8) 
to be given in spherical coordinates $(r,\theta,\varphi)$ by the scalar potential 
\be \phi=  V \left( r+\frac{a^3}{2 r^2}\right)\cos\theta \lb{phi-S} \ee 
where, once again, the zenith for measurement of the polar angle $\theta$ is the positive streamwise 
direction. This implies the potential flow velocity 
\be u_{\phi r}=V\left(1-\frac{a^3}{r^3}\right) \cos\theta, \quad
      u_{\phi \theta} = -V\left(1+\frac{a^3}{2r^3}\right) \sin\theta \lb{u-phi-S} \ee 
whose streamlines are plotted in Figure \ref{fig3} for an axial plane at fixed azimuthal angle $\varphi=0$. 
It is even more useful to represent this flow by the {\it Stokes stream function}  
(\cite{batchelor2000introduction}, \S 2.2),     
given both in spherical coordinates and in cylindrical coordinates $(\sigma,\varphi,x)$ as 
\begin{eqnarray} 
\psi &=&\frac{1}{2} V r^2 \sin^2\theta \left(1-\frac{a^3}{r^3}\right) \cr 
      &=&\frac{1}{2} V \sigma^2 \left[1-\frac{a^3}{(\sigma^2+x^2)^{3/2}}\right] 
\lb{psi-S} \end{eqnarray} 
(\cite{batchelor2000introduction}, \S 6.8) so that 
\be u_{\phi x}=\frac{1}{\sigma}\frac{\partial\psi}{\partial\sigma}, \quad
      u_{\phi \sigma}=-\frac{1}{\sigma}\frac{\partial\psi}{\partial x}. \lb{u-phi-S} \ee 
Here we follow the fluid mechanics literature in denoting $\sigma=\sqrt{y^2+z^2}$ 
to avoid confusion with mass density $\rho.$ Note that streamlines are 
uniquely identified by the values of stream function $\psi$ and azimuthal angle $\varphi.$

Although not needed for the Josephson-Anderson relation, it is worth recalling that the 
pressure $p_\phi$ for the ideal Euler solution 
$\bu_\phi$ follows from the Bernoulli equation
\eqref{bernoulli} as 
\be p_\phi=p_\infty +\frac{1}{4}{\rm V}^2\left(1-\frac{5a^3}{4r^3}\right)\frac{a^3}{r^3} 
+\frac{3}{4}{\rm V}^2\left(1-\frac{a^3}{4r^3}\right)\frac{a^3}{r^3}\cos(2\theta)  
\lb{p-phi-S} \ee 
and, in particular, on the surface of the sphere: 
\be p_\phi(a,\theta) = p_\infty -\frac{1}{6}{\rm V}^2+\frac{9}{16}{\rm V}^2\cos(2\theta).   \ee
This ideal pressure distribution is perfectly symmetrical, with maximum value equal to 
$p_\infty+\frac{1}{2}{\rm V}^2,$ the stagnation pressure, at $\theta=0,\pi$ and 
minimum $p_\infty-\frac{1}{16}{\rm V}^2$ at $\theta=\pi/2.$ This symmetry of $p_\phi$ 
around $\theta=\pi/2$ explains, of course, the vanishing drag for ideal flow past a sphere. 

We can now use the above results to develop a more concrete expression for the 
Josephson-Anderson relation \eqref{dJA-bd} in the flow around a sphere. First,
it is useful to recall that the Stokes stream function is defined for any axisymmetric flow 
$\bu$ so that $d\psi\,d\varphi=\bu\bdot d\bA$ along the streamline labelled by $\psi,$ $\varphi,$
(see \cite{batchelor2000introduction}, \S 2.2). Thus, the element of mass flux appearing 
in \eqref{dJA-bd} is 
\be dJ= \rho\bu_\phi \bdot d\bA=\rho \, d\psi\,d\varphi. \ee 
It is also straightforward to obtain from \eqref{psi-S} an explicit parameterization 
of the streamlines with fixed $\psi$ in the form $x=\pm x(\sigma;\psi)$ for 
$\sigma_{\min}(\psi)<\sigma<\sigma_{\max}(\psi)$ and with azimuthal angle
$\varphi$ constant independent of $\sigma.$ However, for the qualitative arguments 
that we make in the next section, the plots of the streamlines in Figure \ref{fig3} 
are more useful than these analytical expressions. 

We shall furthermore require in the next section a projected version of the vorticity
conservation equation \eqref{vort-cons} for the azimuthal vorticity $\omega_\varphi,$
which we will see is the most crucial component of vorticity for origin of drag. 
Here we follow a general idea of Huggins \cite{huggins1994vortex}, who observed
that one can obtain a balance in any plane for the out-of-plane vorticity $\omega_n$ by 
dotting  \eqref{vort-cons} on the right with the unit vector $\hat{{\bf N}}$ normal to the plane. 
The application of this idea to $\omega_\varphi$ involves one subtlety, however: the 
sets of constant $\varphi$ are half-planes terminating on the $x$-axis, not full planes.
Thus, dotting \eqref{vort-cons} with the unit vector $\hat{{\bf N}}=\hat{\barphi}$ for one such 
half-plane at constant $\varphi$ yields a planar conservation law
\be \partial_t\omega_n+ \grad\bdot \bj_n=0 \lb{om-varphi-bal} \ee 
in the entire plane normal to $\hat{{\bf N}},$ with 
\be \omega_n=\bomega\bdot\hat{{\bf N}}, \quad \bj_n=\bSigma\bdot \hat{{\bf N}}.  \ee
Note that ${\rm j}_{n\,k}=\epsilon_{kn l}(\bu\btimes\bomega-\nu\grad\btimes\bomega)_l$ lies in the  
plane. However, only the upper part of this plane corresponds to azimuthal angle $\varphi,$ 
whereas the lower part corresponds instead to the half plane with azimuthal angle $\varphi+\pi$ 
and whose normal vector is $\hat{\barphi}=-\hat{{\bf N}}.$ Thus, only in the upper half plane does 
$\omega_n=\omega_\varphi,$ whereas in the lower half-plane $\omega_n=-\omega_\varphi.$
We shall see that the conservation
laws \eqref{om-varphi-bal}, which hold separately for each plane through the $x$-axis, are 
very useful for elucidating the physics. 

\subsection{Physical Consequences}

\begin{figure}
\center 
\includegraphics[width=.52\textwidth]{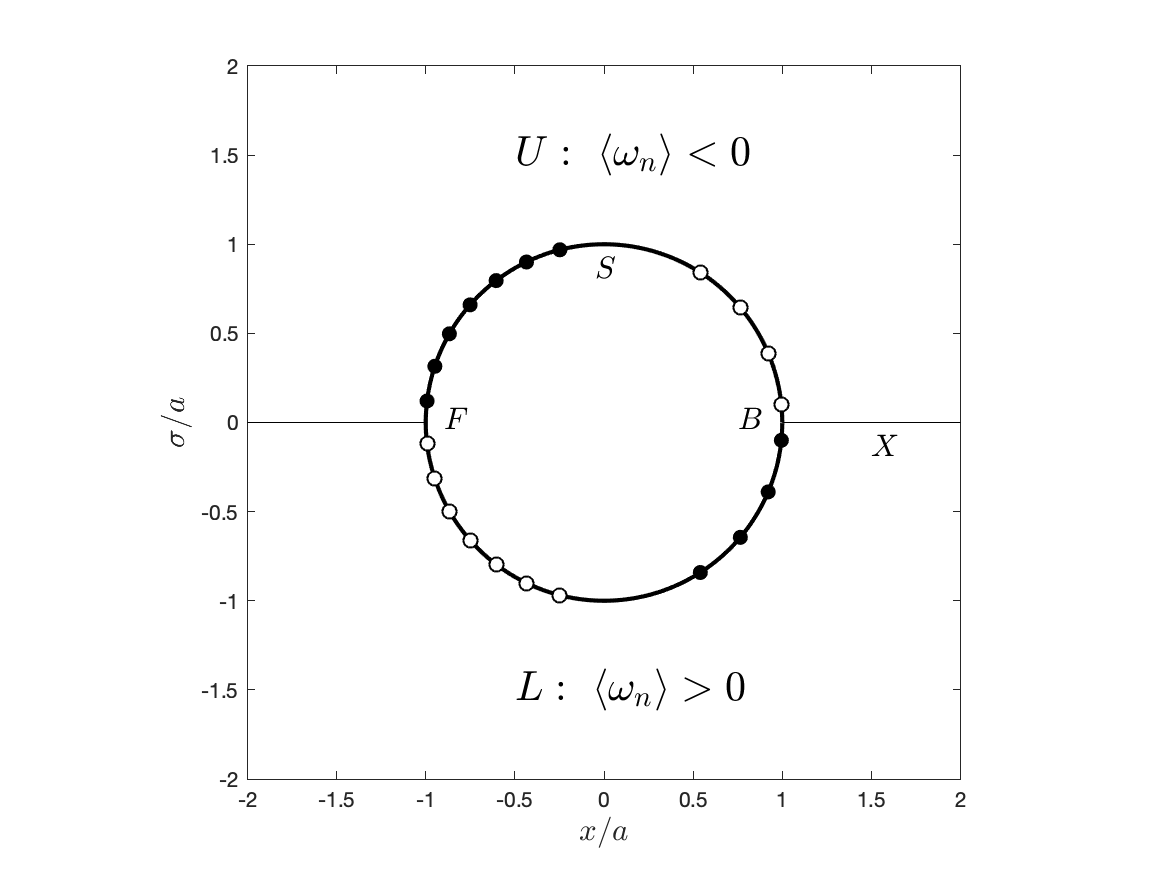}
\caption{Schematic of azimuthal vorticity generation on the surface of the sphere. Following
the convention of Huggins \cite{huggins1994vortex}, we  use white circles to denote 
vorticity out of the plane ($\omega_n>0$)  and black circles to denote vorticity 
into the plane ($\omega_n<0$). The mean normal (azimuthal) vorticity in the upper
 half-plane $U$ is negative and the mean normal (anti-azimuthal) vorticity in the lower
 half-plane $L $ is positive, implying a pole-to-pole asymmetry in the pressure distribution 
 on the surface $S$ of the sphere. The pressure \\ drop along the surface of the sphere
 from $F$ to $B$ is exactly compensated by the pressure rise from $B$ to infinity 
 along the direction of the positive streamwise axis $X$.} 
\lb{fig4} \end{figure}

We shall now exploit the Josephson-Anderson relation \eqref{dJA-bd} and its time-averaged 
form \eqref{meanJA2} in order to develop an exact but qualitative picture of the origin 
of drag in terms of vortex dynamics. Although our picture has the nature of a ``cartoon''
which ignores many complex details of the flow in its different regimes (i)-(viii), we argue 
that it describes the essence of the phenomenon. The concrete predictions that we make 
should be verifiable empirically in each regime, realized somewhat differently by the specific 
flow features that are characteristic of that regime which determine the drag quantitatively. 

An important general deduction from \eqref{dJA-bd} is that vorticity $\bomega\parallel \bu_\phi$
does not contribute directly to drag, and, furthermore, that vorticity flux in the directions of $\bu_\phi$
or $\bomega$ do not contribute. As can be seen from Fig.\ref{fig3}, $\bu_\phi\propto \hat{\bx}$
to a good approximation already at distances about one radius away from the sphere. Thus,
we see that, throughout most of the flow, streamwise vorticity $\omega_x$ makes no direct 
contribution to the drag. Although streamwise vorticity appears in the wake, e.g. in the 
pair of trailing vortices in regime (ii), these features do not contribute anything to \eqref{dJA-bd}.  
On the other hand, at the surface of the sphere where all flow vorticity is generated and in its 
close vicinity, $\bu_\phi\propto\hat{\btheta}$ so that it is the polar vorticity $\omega_\theta$ which does 
not contribute to drag in that region. Since all of the vorticity on the sphere is parallel to its surface, it follows 
that the azimuthal vorticity $\omega_\varphi$ plays the crucial role and, in particular, its viscous 
flux $\Sigma_{r\varphi}$ radially outward. This is our first essential conclusion about the origin of drag 
in the flow past a sphere. 

The next implication of \eqref{dJA-bd} is that outward flux of negative azimuthal vorticity $\omega_\varphi<0$
increases drag, whereas flux of positive azimuthal vorticity $\omega_\varphi>0$ in fact 
{\it decreases} drag. Since the mean power consumption by drag can only be positive 
(see \eqref{meanJA}), it follows that outward flux of negative azimuthal vorticity must be 
larger in magnitude. In other words, there must be an asymmetry in sign of the 
azimuthal vorticity on the sphere, with more area and/or stronger magnitudes where 
$\omega_\varphi<0$ and less areas or magnitudes for $\omega_\varphi>0.$ 
This situation is illustrated by the cartoon in  Figure \ref{fig4}. According to 
Lighthill's theory \cite{lighthill1963introduction}, the rate of generation of $\omega_\varphi$
is exactly $\sigma_\varphi=\frac{1}{a}\frac{\partial p}{\partial\theta},$ so that 
negative azimuthal vorticity is generated by negative or ``favorable'' pressure gradient,
and positive vorticity by positive or ``adverse'' pressure gradient. We thus conclude 
that there must be greater area or greater magnitudes of favorable pressure gradient 
near the front of the sphere than the area or magnitudes of adverse pressure gradient 
toward the back. This results in a pressure asymmetry unlike that for ideal flow,
with the base pressure $p_B$ behind the sphere not fully recovering from its drop in the front, 
thus remaining lower than the stagnation pressure $p_F=p_\infty+\frac{1}{2}{\rm V}^2$ in the front. 
Along the sphere surface $S$
\be p_B-p_F=p(a,0)-p(a,\pi)=\int_\pi^0 \frac{\partial p}{\partial\theta}\,d\theta<0. \ee 
Of course, these conclusions are all in agreement with common observations but it is now 
seen how they are required for the generation of drag by vortex dynamics. 

The picture in Figure \ref{fig4} for an axial plane corresponds in three dimensions to generation 
of azimuthal vortex loops or rings on the surface of the sphere.  Those in the front of the sphere 
have $\omega_\varphi<0$ while those toward the rear have mostly $\omega_\varphi>0.$
 As these rings flow radially outward from the surface, they encircle greater amounts 
 of the potential mass-flux $J,$ with the negative-$\omega_\varphi$ rings removing 
 proportionate energy from the potential flow and positive-$\omega_\varphi$ rings
 returning that energy. Since $\omega_\varphi<0$ predominates, the net effect is 
 a transfer of kinetic energy from potential to vortical motions, where it is ultimately dissipated 
 by viscosity.  In the axisymmetric wake regime (i) this picture is exact, because all
 vorticity is azimuthal. In higher-$Re$ regimes there will also be some polar vorticity 
 $\omega_\theta$ generated on the sphere by azimuthal pressure gradients as 
 $\sigma_\theta=-\frac{1}{a}\frac{\partial p}{\partial\varphi},$ but these components 
 and their radial flux from the sphere do not contribute to the drag. See Figure \ref{fig5} 
for a rough cartoon of the mechanism.  It should be noted that this entire picture developed 
on the basis of the relation \eqref{dJA-bd} is consistent with the direct formula \eqref{F-vort} for the drag 
 force ${\bf F}$, since the asymmetry  in pressure $P$ and the negative azimuthal vorticity 
 $\bomega$ on the surface of the sphere result in pressure and viscous forces both opposing 
 the fluid velocity $\bV.$

\begin{figure}
\center
\includegraphics[width=.4\textwidth]{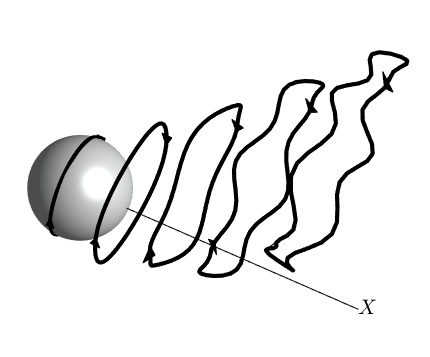}
\caption{Rough cartoon of the drag mechanism by generation and outward flux of vortex loops
with negative azimuthal vorticity. The rings are generated by pressure drop along the surface.  
As the rings expand outward, they enclose greater mass flux $J$ and subtract proportionate 
energy from the potential flow. Although the rings initially loop around the streamwise axis $X,$ 
they drift across in time. This implies a flux of azimuthal vorticity across $X$ and allows the 
pressure to recover far downstream from its drop around the sphere.} 
\lb{fig5} \end{figure}

As the vortex loops grow, they enter the region where $\bu_\phi\propto \hat{\bx}$ 
and then azimuthal vorticity $\omega_\varphi$ is no longer the only relevant component.
Indeed, referring now to the cylindrical coordinate system, axial vorticity $\omega_\sigma$ 
appears in this region through the tilting of azimuthal vorticity by the shear in the wake, e.g. 
as alternating hairpin vortices observed in the shedding regime (iii). It is easy to check 
that the flux $\Sigma_{\varphi\sigma}$ 
due to opposite azimuthal motion of the two legs of growing hairpin vortices, one with
$\omega_\sigma>0$ and the other with $\omega_\sigma<0,$ contributes also 
to drag through the relation \eqref{dJA-bd} in this spatial region, in addition to the flux 
$\Sigma_{\sigma\varphi}$ of negative azimuthal vorticity axially outward. In fact, these two 
effects are basically the same because of the anti-symmetry $\Sigma_{\varphi\sigma}
=-\Sigma_{\sigma\varphi}$ and both correspond to growth of the tilted vortex rings,
which encompass an increasing amount of mass flux $J$ and gain a proportionate 
amount of energy from the potential flow. This is the essential mechanism of drag 
in terms of vortex dynamics for the flow around a sphere as already illustrated in Figure \ref{fig5}.  

Conservation of azimuthal vorticity has one very important additional implication 
for this process. We showed at the end of section \ref{sec:gbal} that $\Delta h'=0$
along each streamline of the potential flow. Since now $\dot{\phi}=0,$ 
$h'=h$ and $h(\ell=\pm\infty)=h_\infty=p_\infty+\frac{1}{2}{\rm V}^2$ at each limit of the 
streamline. Notice that this value is the same as the stagnation pressure at the 
front of the sphere $h_F=p_F=p_\infty+\frac{1}{2}{\rm V}^2$ and also that $h_B=p_B$ 
at the base point behind the sphere, since the fluid velocity vanishes everywhere 
on the surface. We thus see that the decrease of $h$ on the surface $S$ of the sphere 
\be \int_\pi^0 \frac{\partial h}{\partial \theta}(a,\theta)\, d\theta = h_B-h_F<0 \ee 
is exactly equal and opposite to the increase of $h$ along the streamwise axis $X$ from the base 
of the sphere to infinity:
\be \int_a^\infty \frac{\partial h}{\partial x}(0,x)\, dx=h_\infty-h_B>0  \ee 
See Figure \ref{fig4}.  
Since $\langle\Sigma_{r\varphi}\rangle=\frac{1}{a}\frac{\partial \langle h\rangle}{\partial\theta}$
on $S$ and $\langle\Sigma_{\sigma\varphi}\rangle=\frac{\partial \langle h\rangle}{\partial x}$
on $X$, it follows that, on average, the net negative azimuthal vorticity generated 
on the surface of the sphere is exactly cancelled by flux of opposite-signed vorticity 
across the $x$-axis. This fact was previously noted by Brown \& Roshko \cite{brown2012turbulent}
for flow around a cylinder and by Terrington et al. \cite{terrington2021generation} for flow around 
a sphere. This exact balance is what permits a steady state to exist with a mean negative 
value $\langle\omega_\varphi\rangle<0$ in the axial half-planes for each fixed value of $\varphi.$
The precise origin of the flux across the $x$-axis is still debated, but it is most likely 
due to lateral motion of shed vortex rings, which initially loop around the $x$-axis but 
drift across the $x$-axis to become unlinked from it as they advect downstream. See Figure \ref{fig5}. 
Although this is a 
``cartoon picture'' assuming simple closed vortex rings, it presumably has an exact 
counterpart in the actual vortex motions, because solenoidality requires a pairing 
of the points with  $\omega_\varphi<0$ on opposite sides of the $x$-axis and 
viscous or turbulent diffusion will tend to mix those points from one side to the other. 

In this section we have attempted to give a clear and intuitive description of the origin 
of drag in terms of a schematic picture of the vortex motions. Although qualitative, 
this picture makes specific predictions in terms of the signs of the vorticity components 
$\omega_j$ and their fluxes $\Sigma_{ij},$ which can be checked empirically by measuring 
these quantities in numerical simulation or experiment and evaluating their contribution to the 
drag through the relation \eqref{dJA-bd}. We note that the numerical 
study \cite{terrington2021generation} has already used Huggins' vorticity flux 
\eqref{Sigma} to illuminate other aspects of the vortex dynamics in the wake behind 
a sphere, such as tilting of the azimuthal vorticity in the streamwise direction 
\footnote{Note that the authors of \cite{terrington2021generation} refer to Huggins' 
flux in \eqref{Sigma}, as ``Lyman-Huggins flux'', citing also \cite{lyman1990vorticity}. 
This double attribution is not really proper, in our opinion, since Lyman considered vorticity generated 
only at the boundary, whereas it was Huggins who first considered this flux in the interior of the flow.}

\subsection{Superfluid Comparisons} 

The cartoon picture that we presented in the previous section in order to interpret
the exact Josphson-Anderson relation \eqref{dJA-bd} should be more literally 
correct for a superfluid, where vorticity is quantized and vortex lines are discrete objects 
whose motion is objectively defined. In fact, the most fundamental 
differences between our theoretical results and those for superfluids arise not 
from the differences between classical and quantum fluids, but instead from the 
differences between incompressible and compressible. In superfluids there 
are generally substantial compressibility effects, due to emission of phonons
that propagate at the finite speed of sound. Of course, the Josephson-Anderson 
relation arose in the study of superfluids and its application there to understanding 
drag and critical velocities is well-established. It is therefore worth reviewing briefly the 
existing literature in order to point out both the similarities and the differences 
with classical incompressible fluids. 

The pioneer work concerning drag acting on bodies moving through a superfluid is that 
of Frisch et al. \cite{frisch1992transition}, who adopted the zero-temperature 
Gross-Pitaevskii model to study numerically the motion of a disk at velocity ${\rm V}$ through 
a 2D superfluid. We remark that it is straightforward to extend our own analysis for incompressible 
classical fluids to 2D. Already showing the important effects of compressibility,
the authors of  \cite{frisch1992transition} identified the critical velocity ${\rm V}_c$ for appearance 
of drag to be that for which the {\it local} velocity $u$ on the surface of the disk exceeds the 
sound speed $c_s.$ At this velocity ${\rm V}_c\doteq 0.44\,c_s$, quantized vortices are 
nucleated and emitted as a wake behind the disk. Although compressibility plays a crucial 
role in their generation, the vortices themselves are incompressible excitations in the superfluid.
The picture proposed in \cite{frisch1992transition} has been further developed in many 
subsequent works on this same problem 
\cite{jackson1998vortex,winiecki1999pressure,winiecki2000vortex,
winiecki2001numerical,huepe2000scaling,nore2000subcritical,sasaki2010benard}.
In particular, we note that the generation of the vortices has been verified 
to occur by the $2\pi$-phase slip mechanism \cite{jackson1998vortex} 
and their shedding occurs with the Josephson frequency corresponding 
to the difference in the generalized chemical potential $\mu_T=g n+\frac{1}{2}m u^2$
that develops between the exterior flow and the wake behind the disk with low density 
$n$ and low speed $u$ \cite{winiecki1999pressure}. Thus, at least for 
${\rm V}<c_s,$ there is great similarity with the theory that we have 
developed for classical, incompressible fluids. Of course, for supersonic motions  
with ${\rm V}>c_s,$ new compressible effects can be observed in superfluids, such 
as drag by phonon radiation and standing bow waves in front of the disk 
\cite{winiecki2000vortex,winiecki2001numerical}, which have no 
parallel in the incompressible theory developed here. 

Studies of superfluid drag have since been extended to 3D, with the superfluid 
modeled again by Gross-Pitaevskii and the moving object by a suitable time-dependent 
potential \cite{winiecki2000motion,stagg2014quantum}. The object was taken in 
\cite{winiecki2000motion} to be spherical and subject to constant force ${\bf F},$ 
whereas \cite{stagg2014quantum} considered more general ellipsoidal bodies and 
moving at constant velocity $\bV.$ The picture emerging from these 3D simulation 
studies is even more strikingly similar to the one that we have derived for classical
fluids. In both studies, quantized vortex rings are excited at the surface of the object 
when it reaches the critical speed ${\rm V}_c,$ with the vorticity oriented in the 
negative azimuthal direction (according to our coordinate conventions). In the 
case of the body moving at constant speed studied in \cite{stagg2014quantum},
the ring vortices are shed into the wake where they grow, extracting energy 
from the potential flow, and also drift cross-stream so that the asymmetry in 
the vortex polarity is relaxed far downstream. The observed simulation results are 
very close to our sketch in Figure \ref{fig5}. The case with constant applied force 
${\bf F}$ studied in \cite{winiecki2000motion} shows a bit more complex 
behavior, because the body decelerates when the vortex ring is emitted.
At lower forcing, the quantized vortex ring reattaches to the spherical body
and remains pinned there as an arch, legs perpendicular to the surface, even 
as it continues to grow and expand outward. This regime has no strict classical 
analogue, although it distantly resembles the reattachment of the separating 
laminar boundary layer observed in the ``drag crisis'' of the classical regime (vi). 
At higher forcing, however, the vortex rings completely detach and are shed 
in the wake, again very similar to our Figure \ref{fig5}. 

Somewhat ironically, the detailed Josephson-Anderson relation was derived 
by Huggins \cite{huggins1970energy} assuming flow incompressibility, and we are 
aware of no full extension to the Gross-Pitaevskii model of a superfluid. The original 
Josephson-Anderson phase relation \eqref{JArelation} is, of course, directly 
embodied in the Gross-Pitaevskii equation (with an additional ``quantum pressure'' term),
but this implies no direct connection of vortex motion with energy dissipation. 
Thus, based on our results in the present paper, we currently have a better understanding 
for a classical incompressible fluid how energy dissipation is associated to vorticity flux  
than we do for quantum superfluids, where the Josephson-Anderson relation originated! 
 
\section{Why is It Important?} 

To briefly summarize our results in this paper, we have derived the detailed 
Josephson-Anderson relation \eqref{dJA-bd} for incompressible fluid flow around a finite solid body, 
relating drag and energy dissipation to vorticity flux and implying a time-averaged 
version \eqref{meanJA} valid for the local steady-states of the fluid wake. We 
have furthermore discussed in detail the origins of drag in terms of vortex motion
for the concrete example of flow past a sphere, obtaining numerous predictions that 
can be checked empirically. But why are these results important? 

We believe first of all that the Josephson-Anderson relation is important because 
of the theoretical unity that it brings to our understanding of drag for both quantum 
and classical fluid systems, across all Reynolds number ranges of the latter. 
The relation clearly identifies what is essential for drag in the vortex dynamics and 
statistics, bypassing all details of secondary importance. In the 
classical case, these ``details'' include fascinating phenomena such as boundary-layer separation, 
formation of streamwise vortices by tilting, transition to turbulence first in the wake 
and then in the boundary-layer, etc. All such ``details'' are, of course, crucial in order
to determine the drag quantitatively, but only insofar as they modify the primary process:
flux of vorticity across the potential mass flux. The precise vortex dynamics and statistics 
which contribute to the Josephson-Anderson relation must be explained in any quantitative 
theory of drag. 

The relation \eqref{dJA-bd} furthermore sheds new insight on the D'Alembert 
paradox \cite{dalembert1749theoria,grimberg2008genesis}. We should emphasize that,
in our opinion, the paradox in its original form had its solution in the realization by Saint-Venant  
of the importance of even a very small viscosity \cite{saintvenant1846resistance}, as further 
elaborated by Prandtl \cite{prandtl1904motion} and others \cite{stewartson1981dalembert}.  However, 
the paradox has recently arisen in a new guise, because of two apparently conflicting facts, one 
empirical and the other mathematical. The empirical fact from laboratory observations is that the 
dimensionless drag coefficient $C_D(Re)$ for flow around a finite object such as a sphere 
apparently tends to a non-zero constant value as $Re\to\infty$ \cite{achenbach1972experiments}. \\ 
The mathematical  fact is that, under very modest assumptions, the viscous Navier-Stokes solution 
must tend in the limit $Re\to\infty$ (along a suitable subsequence) to a weak solution of the 
incompressible Euler equations \cite{drivas2019remarks}. The ``paradox'' is then how the 
limiting weak Euler solution can produce a non-vanishing drag. 

This puzzle is obviously connected with the Onsager theory of turbulence in terms 
of dissipative weak solutions of the Euler equations  \cite{onsager1949statistical,
eyink2006onsager,eyink2018review}.  In fact, our results in this paper open up 
the possibility of a novel analysis that directly relates the D'Alembert paradox 
with Onsager's theory. One crucial observation is that the new potential-vortical formulation 
\eqref{NS-omega-mom} of incompressible Navier-Stokes is a local conservation equation for 
$\rho \bu_\omega,$ the density of ``vortex momentum'', and it thus has a weak interpretation. 
We may therefore apply in this formulation the recent techniques to derive Onsager's theory for 
wall-bounded flows \cite{bardos2018onsager,drivas2018onsager,bardos2019onsager,chen2020kato},
obtaining necessary conditions for non-vanishing drag and dissipation in the infinite Reynolds-number limit.  
The relevant limiting Euler solutions must obviously describe rotational flow very distinct 
from the smooth potential-flow solution.  The local energy balances \eqref{Eom-loc}, \eqref{Eint-loc} 
make sense also in a weak interpretation and, in the infinite-$Re$ limit, the viscous dissipation appearing 
in \eqref{Eom-loc} will yield a ``dissipative anomaly'' like that appearing in the work of Duchon \& Robert 
for periodic domains \cite{duchon2000inertial}. However, {\it a priori} the energy transfer term in 
the two equations \eqref{Eom-loc}, \eqref{Eint-loc} may have no limit because the pointwise
product $\bu_\omega\btimes \omega$ has no clear meaning when $\bu_\omega$ and $\bomega$
for the limiting Euler solution are merely distributions.  Fortunately, this term can be rewritten as a space-gradient 
using the familiar vector calculus identity
\be \bu_\omega\btimes\bomega=-\grad\bdot\left[\bu_\omega\bu_\omega-\frac{1}{2}|\bu_\omega|^2
{\bf I} \right] \ee 
and this is a well-defined distribution even if $\bu_\omega$ is only square-integrable. 
Because the potential-flow solution $\bu_\phi$ is smooth, this observation allows us to rewrite 
the transfer term in the Josephson-Anderson relation \eqref{dJA-bd} as 
\be \int_\Omega \bu_\phi\bdot(\bu_\omega\btimes\bomega) \, dV
= \int_\Omega \grad\bu_\phi\bdots\bu_\omega\bu_\omega\, dV \ee 
and, in this form, the relation can be valid even for the limiting weak solutions of Euler 
equations \cite{eyink2021onsager}. It is very natural that drag for the limiting 
Euler solution should be connected with vorticity flux since, as emphasized by 
Morton  \cite{morton1984generation}, generation of vorticity at the wall by tangential
pressure-gradients is a purely inviscid process. 

Of more immediate importance for practical applications is the new insight that the 
Josephson-Anderson relation \eqref{dJA-bd} provides on techniques 
for drag reduction. As discussed previously \cite{huggins1994vortex,eyink2008turbulent},
a ``drag problem'' occurs not only in turbulent flow at high Reynolds number 
but also in high-temperature superconductors because, above a critical electric current,
quantized magnetic flux-lines are nucleated which migrate cross-current and create 
a voltage drop and effective resistance \cite{josephson1965potential,bishop1993resistance}.  
The technological solution that has been found is to introduce some sort of disorder to pin the 
quantized lines so that they are not free to cross the current, permitting resistance-less 
conduction up to higher values of electric current. It might be thought that vortices cannot 
be so easily ``pinned" in a classical fluid with smoothly distributed vorticity, but, in fact, the 
Josephson-Anderson relation \eqref{dJA-bd} tells us that any mechanism which 
reduces drag must somehow decrease vorticity flux across the potential mass current! 
This includes mechanisms such as drag reduction by polymer additives or the Toms effect,
\cite{toms1948some,white2008mechanics}, whose efficacy is well-documented but whose 
detailed physical explanation is still debated. We believe that empirical investigation of 
the vorticity flux will provide new clues into the underlying mechanism. 
 
 Granted that the detailed Josephson-Anderson relation \eqref{dJA-bd} has important 
 implications, it then becomes interesting to explore the full range of its validity. Although
 bodies of finite extent are most realistic, it should be illuminating to generalize the relation 
 to more idealized geometries such as flow past cylinders and semi-infinite plates. Validity of 
 the relation for compressible flow would greatly widen the range of applications and we believe
that this could be possible, especially for barotropic models where smooth Euler solutions satisfy a 
Kelvin circulation theorem and potential flow conditions are preserved in time (including 
the Gross-Pitaevskii model). Finally, there should be a connection with stochastic Lagrangian
representation of the vorticity dynamics \cite{constantin2008stochastic,constantin2011stochastic, 
eyink2020Astochastic,eyink2020Bstochastic}, which gives a more precise meaning to the ``motion'' 
of vortex lines in a classical fluid. This approach has recently been generalized to solve the Helmholtz 
equation with the Lighthill vorticity source as Neumann boundary conditions \cite{eyink2021stochastic,
wang2021origin}, but it is not yet clear how to relate the stochastic Lagrangian trajectories 
to the Huggins vorticity flux tensor in the flow interior. These stochastic representations are 
an exact mathematical approach to realize Huggins' suggestion \cite{huggins1971dynamical} 
of a probability interpretation of the vorticity field.  

\acknowledgements 
We wish to thank Ping Ao for his insistence to us in 2006 that energy dissipation in 
classical fluids should be explainable by the Josephson-Anderson relation, which 
led us to discover the works of Huggins. We thank also the Simons Foundation for support 
of this work through Targeted Grant in MPS-663054, ``Revisiting the Turbulence Problem 
Using Statistical Mechanics.''

\appendix

\section{Derivation of Huggins' Relation}\lb{app:huggins}  

For completeness, we reproduce here the derivation of the detailed Josephson-Anderson relation
for the channel-flow geometry, originally obtained by Huggins. The starting point is the Bernoulli
equation for the scalar potential of the ideal flow: 
\be \dot{\phi}+\frac{1}{2}|\bu_\phi|^2 +p_\phi+U=0.  \lb{bernoulli-H} \ee 
Its space-gradient is the Euler equation for $\bu_\phi=\grad\phi,$ which, 
subtracted from the Navier-Stokes equation \eqref{NS-ext}, yields the equation of motion 
for $\bu_\omega:=\bu-\bu_\phi: $
\be \partial_t\bu_\omega=\bu\btimes\bomega-\nu\grad\btimes\bomega -\bg
-\grad\left(p_\omega+\frac{1}{2}|\bu_\omega|^2+\bu_\omega\bdot\bu_\phi\right), \lb{uom-H} \ee 
with the boundary conditions
\be \left.\bu_\omega\right|_{S_{in}}=\bzed, \quad \left.\bu_\omega\right|_{S_{out}}=\bzed, 
\quad \left.\bu_\omega\right|_{S_{w}}=\bzed. \ee
Dotting \eqref{uom-H} with $\bu_\omega$ and integrating over the channel domain $\Omega$ 
immediately yields 
\be \frac{dE_\omega}{dt} = {\mathcal T} -\int_\Omega [\eta|\bomega|^2+\rho\bu\bdot\bg] \,dV
\lb{Eomega-Huggins} \ee
which is \eqref{Eomega-H} in the main text. On the other hand, the total kinetic energy
$E(t)=(1/2)\int\rho|\bu|^2\,dV$ satisfies 
\be \frac{dE}{dt} = \int dJ (h_{in}-h_{out})-\int_\Omega [\eta|\bomega|^2+\rho\bu\bdot\bg] \, dV
\lb{E-Huggins} \ee
and since $E=E_\phi+E_\omega$ by the Kelvin minimum energy theorem, we obtain 
\be \frac{dE_\phi}{dt} = \int dJ (h_{in}-h_{out}) - {\mathcal T} \lb{Ephi-Huggins}  \ee
which is \eqref{Ephi-H} in the main text. Finally, 
\begin{eqnarray}  
\frac{dE_\phi}{dt} &=&\int_\Omega \rho \bu_\phi\bdot\dot{\bu}_\phi \,dV\cr 
&=& 
 \int_\Omega \rho \grad\bdot(\dot{\phi}\bu_\phi)\,dV 
 = \int dJ (\dot{\phi}_{out}-\dot{\phi}_{in}) 
\lb{Ephi-dot-Huggins}  \end{eqnarray}  
by using the divergence theorem and $dJ=\rho \bu_\phi\bdot d\bA.$ Combining 
\eqref{Ephi-Huggins} and \eqref{Ephi-dot-Huggins} yields 
\be {\mathcal T}= \int dJ (h_{in}'-h_{out}') \lb{dJA-ch-Huggins} \ee
which is the ``detailed Josephson relation'' \eqref{dJA-ch} first derived by Huggins. 

\section{Vortex Momentum and Impulse}\lb{app:vortmom}

We give here another derivation of \eqref{cantwell}, different from that of \cite{cantwell1986viscous}.  
Using the identity 
\be \bx\btimes \bomega= x_i\grad u_{\omega i}-(\bx\bdot\grad)\bu_\omega \ee 
we find after integration by parts that 
\begin{eqnarray}
&& \int_\Omega\bx\btimes\bomega\, dV = \int_\Omega (-1+3)\bu_\omega\,dV \cr 
&& -\int_{\partial B} \bx\btimes(\hat{\bn}\btimes\bu_\omega)\, dA 
+\lim_{R\to\infty}\int_{S_R} \bx\btimes(\hat{\br}\btimes\bu_\omega)\, dA \cr
&& 
\end{eqnarray}   
or from the definition of $\bI_\omega$
\be 2\bI_\omega = 2\int_\Omega \bu_\omega\, dV +\lim_{R\to\infty}\int_{S_R} \bx\btimes(\hat{\br}\btimes\bu_\omega)\, dA \ee 
Using the asymptotic far-field expansion of $\bu_\omega$ 
\begin{eqnarray}
&& \lim_{R\to\infty}\int_{S_R} \bx\btimes(\hat{\br}\btimes\bu_\omega)\, dA 
\ =\ \lim_{R\to\infty}\int_{S_R} R\hat{\br} \btimes \frac{\hat{\br}\btimes(-\bI_\omega)}{4\pi R^3}\ R^2\,d\Omega  \cr
&& \hspace{50pt}  =\frac{1}{4\pi} \int (\bI_\omega-(\bI_\omega\bdot\hat{\br})\hat{\br} )\ d\Omega 
\ = \ \frac{2}{3}\bI_\omega
\end{eqnarray} 
Thus, we finally obtain 
\be \int_\Omega \bu_\omega\, dV = \frac{2}{3}\bI_\omega. \ee

\bibliography{bibliography.bib}

\end{document}